\documentclass[aps,physrev,preprint,groupedaddress]{revtex4-2}

\usepackage{graphicx}%
\usepackage{multirow}%
\usepackage{amsmath,amsfonts}%
\usepackage{amssymb}
\usepackage{amsthm}%
\usepackage{mathrsfs}%
\usepackage[title]{appendix}%
\usepackage{xcolor}%
\usepackage{textcomp}%
\usepackage{booktabs}%
\usepackage{algorithm}%
\usepackage{algorithmicx}%
\usepackage{algpseudocode}%
\usepackage{listings}%
\usepackage{physics}%
\usepackage{tikz, orcidlink}%
\usepackage{array}
\usepackage{multirow}

\newcommand{\bs}{\boldsymbol}
\definecolor{revisioncyan}{RGB}{0,160,180}

\newcommand{\Id}{\mathbb I}
\newcommand{\exd}{\mathrm d}
\newcommand{\OmR}{\Omega_{\mathrm R}}
\newcommand{\Argop}{\operatorname{Arg}}
\newcommand{\Rotz}{\mathcal R_z}
\newcommand{\rss}{r_{\mathrm{ss}}}
\newcommand{\tss}{\theta_{\mathrm{ss}}}
\newcommand{\req}{r_{\mathrm{eq}}}
\newcommand{\nth}{n_{\mathrm{th}}}

\newcommand{\Geff}{\Gamma}

\begin{document}

\title{Geometric Phases of a Driven Qubit: Comparing Berry and Uhlmann Holonomies}

\author{Hyeonseok Yang$^{1,2}$}
\email[]{yhs1802@gmail.com}
\author{Changsuk Noh$^1$}
\email[]{cnoh@knu.ac.kr}

\affiliation{$^1$Department of Physics, Kyungpook National University, Daegu, South Korea}
\affiliation{$^2$Department of Chemistry, Graduate School of Science, Kyoto University, Kyoto, 606-8502, Japan}

\date{\today}

\begin{abstract}
The Berry phase arises naturally from the adiabatic dynamics of a pure quantum state, whereas the Uhlmann phase is usually formulated kinematically for a prescribed path of mixed states. For a qubit subject to a uniformly gapped conical drive, we obtain the Uhlmann connection and holonomy in closed form for the equilibrium Gibbs cycle. We then solve the corresponding Lindblad equation exactly in the rotating frame and show that its steady state forms a nonequilibrium limit cycle that lags the instantaneous Gibbs cycle. As the driving becomes slow, the Uhlmann phase approaches the equilibrium Uhlmann phase, which cooling then reduces to the Berry phase: in the joint adiabatic and low-temperature limit, the dynamical Uhlmann phase therefore converges to the Berry phase. This grounds the Uhlmann--Berry correspondence in the physical dynamics of an open system. We also characterize the finite-driving corrections, together with the discontinuous $\pi$ jump that the Uhlmann phase undergoes on the equatorial cycle as the temperature is varied. Finally, we examine an isolated transversal gap closing, where this zero-temperature correspondence can break down: the pure-state path becomes open, its closure is ambiguous, and gap-opening regularizations close it along a geodesic set by the direction of the bias field, splitting the Berry phase into a one-parameter family of values. The Gibbs path instead closes smoothly through the maximally mixed state, so the Uhlmann holonomy requires no regularization and remains unique and continuous at every finite temperature. In the low-temperature limit the Uhlmann phase selects a single member of the Berry family---the geodesic closure in the osculating plane, fixed by the velocity and acceleration of the driving field. The Uhlmann--Berry correspondence of the gapped regime thus survives the gap closing as a geodesic selection rule.
\end{abstract}

\maketitle

\section{Introduction}

Geometric phases arise when a quantum system undergoes cyclic evolution in parameter space, leading to a phase accumulation that is independent of the dynamical phase and determined solely by the geometry of the path. The prototypical example is the Berry phase, originally formulated for pure states undergoing adiabatic and cyclic evolution in a general parameter space~\cite{Berry1984,Simon1983}. In the special case of a two-level system, it acquires an intuitive interpretation as half the solid angle subtended on the Bloch sphere by the closed path. Aharonov and Anandan later generalized the notion of geometric phase beyond the adiabatic regime to arbitrary cyclic evolutions~\cite{AharonovAnandan1987}. Geometric phases have since been experimentally observed in systems such as photons in optical fibers, neutron spin rotation, and electron interferometry~\cite{TomitaChiao1986, BitterDubbers1987,Tonomura1988new}, and have also been theoretically investigated in concrete quantum-optical settings including two-level atomic systems~\cite{Tewari1989}. More broadly, geometric phases now play an important role in the characterization of topological matter and in robust schemes for quantum information processing~\cite{SjoqvistGPQI2015}.

The conventional formulation of the Berry phase, however, is intrinsically constructed for pure states. Realistic quantum systems at finite temperature or in nonequilibrium environments are generally described by mixed states represented by density matrices, which naturally raises the question of how geometric phases may be extended beyond the pure-state framework. One mathematically rigorous approach is the Uhlmann phase, defined through purification of the density matrix together with a parallel-transport condition~\cite{Uhlmann1986}. The associated Uhlmann parallel transport and Bures metric can be evaluated explicitly in low-dimensional Hilbert spaces~\cite{Hubner1993}. An alternative formulation was proposed by Sj\"{o}qvist \textit{et al.}, who generalized the Mach--Zehnder interferometric setup to unitary evolution of mixed states and introduced the interferometric geometric phase~\cite{Sjoqvist2000}. The existence of multiple formulations of mixed-state geometric phases has led to ongoing efforts to clarify their relations and physical distinctions~\cite{Hou2023comparative,Hou2024QGT}.

A particularly important question concerns the pure-state limit, namely whether the Uhlmann phase reduces to the Berry phase in the zero-temperature limit. Previous studies have reported such Uhlmann--Berry correspondence in several settings, including coherent states, and attempts have been made to formulate it as a general property, modulo certain exceptional cases, with conditional proofs~\cite{WangUBC2023}. However, the Uhlmann process is generally incompatible with the Hamiltonian dynamics governing the density matrix, so that mixed states accumulate dynamical contributions during time evolution~\cite{Guo2020,Guo2020erratum}. In addition, the Uhlmann phase is known to decrease rapidly as the decoherence rate increases under isotropic decoherence~\cite{Tidstrom2003}. These observations suggest that the Uhlmann phase is not merely a mixed-state extension of the Berry phase, but may exhibit qualitatively distinct behavior arising from its non-Abelian structure and coupling to the environment. Indeed, the gauge structure associated with Uhlmann holonomy is generally given by $U(n)$, in contrast to the $U(1)$ structure underlying Berry holonomy~\cite{Uhlmann1986,WangUBC2023}.

Uhlmann holonomy has also been actively studied in the context of finite-temperature topological invariants. Viyuela \textit{et al.} introduced the density-matrix Chern insulator and the topological Uhlmann number to characterize symmetry-protected topological phases at finite temperature~\cite{Rivas2013,Viyuela2014_2D,Viyuela2015SPT}, and the corresponding signatures were later observed experimentally in superconducting qubit systems~\cite{Viyuela2018Obs}. Subsequent studies further extended these ideas to the robustness of topological protection under Lindblad dynamics and Uhlmann--Chern numbers~\cite{He2018ThermalChern,He2022Lindblad}, as well as to finite-temperature phase transitions in spin-$j$ and driven spin systems~\cite{Hou2021spinj,Villavicencio2023,Wang2025spin}. These developments are closely connected to broader studies of topological phenomena in open quantum systems, including Uhlmann curvature in nonequilibrium steady-state transitions~\cite{Carollo2018}, geometric phases along quantum trajectories~\cite{Viotti2023}, and topological order in driven-dissipative Kitaev chains~\cite{vanCaspel2019}.

All of the above concerns loops along which the Hamiltonian remains gapped. When the control field of a two-level system is swept through the degeneracy, the direction $\hat{\boldsymbol{n}}=\boldsymbol{d}/d$ traces an open curve on the Bloch sphere whose endpoints are antipodal, and the geodesic rule for open-path geometric phases~\cite{Pancharatnam1956,SamuelBhandari1988,MukundaSimon1993} becomes ambiguous: infinitely many great-circle arcs close the path, and different closures enclose different solid angles. The associated $\pm\pi$ phase jumps and operational closing prescriptions have been analyzed for two-level systems~\cite{RakhechaWagh1996,GarzaSoto2023} and tested in neutron and atom interferometry~\cite{Wagh1998,Zhou2020}, while on the mixed-state side the Uhlmann--Berry correspondence has recently been reported to break down at spectral degeneracies~\cite{Wang2025degenerate}. A controlled setting in which both holonomies can be followed through one and the same gap closing has, however, been lacking.

In this work we compare Berry and Uhlmann holonomies for a single qubit Hamiltonian with a driving field, $H(t)=-\boldsymbol{d}(t)\cdot\boldsymbol{\sigma}$, in three steps. First, for conical driving, we derive the Uhlmann connection and holonomy in closed form and compare the mixed-state phase with the Berry phase as functions of bath temperature and the control-field parameters (Secs.~II--III). Second, we solve the Lindblad equation under the driven field exactly in the rotating frame. The steady state is a limit cycle whose lag components relative to the instantaneous Gibbs loop scale as powers of $\eta=\omega/\omega_0$, with bath-related parameters. This gives the ordered hierarchy $\gamma_U^{\rm dyn}\to\gamma_U^{\rm eq}\to\gamma_B$: The first limit is adiabatic, reducing the dynamical limit cycle to the equilibrium Gibbs loop. In the second, zero-temperature limit, the Gibbs state becomes the rank-one projector $|g(t)\rangle\langle g(t)|$, and the projected Uhlmann transport reduces to the Berry $U(1)$ holonomy (Sec.~IV). Third, we let the drive close the gap at an isolated instant. The Berry construction then yields a one-parameter family $\gamma_B=-\Omega[C]/2-\alpha$ via regularization, whereas the Uhlmann holonomy remains regularization-independent and continuous through the transverse crossing. In the low-temperature limit the Uhlmann phase selects the osculating-plane value $-\Omega[C]/2$, so the gap-closing correspondence is a selection from the Berry family rather than a one-to-one equivalence (Sec.~\ref{sec:gapclosing}). The final step builds on two geometric results derived in a companion paper: the intrinsic solid angle of an open path with antipodal endpoints and the dependence of the regularized phase on the chosen closure~\cite{YangNoh2026geodesic}.

\section{Berry Holonomy}

\subsection{The Driven Qubit System}

We consider a driven qubit system governed by the time-dependent Hamiltonian as follows
\begin{equation}
H(t)= -\boldsymbol{d}(t)\cdot\boldsymbol{\sigma},
\qquad
\boldsymbol{d}(t)=\bigl(d_0\cos\omega t,\ d_0\sin\omega t,\ d_1\bigr),
\label{eq:H_lab}
\end{equation}
Throughout, we set $\hbar=k_B=1$. Here $\omega=2\pi/T$, $\sigma_j$ with $j\in\{x,y,z\}$ are the Pauli matrices, and $T$ is the driving period. 
The eigenvectors of the system Hamiltonian can be written as
\begin{equation}
|g(t)\rangle=
\begin{pmatrix}
\cos\frac{\theta}{2}\\
e^{i\phi(t)}\sin\frac{\theta}{2}
\end{pmatrix},
\qquad
|e(t)\rangle=
\begin{pmatrix}
-\sin\frac{\theta}{2}\\
e^{i\phi(t)}\cos\frac{\theta}{2}
\end{pmatrix}.
\label{eq:eigenkets_singlephase}
\end{equation}
with eigenvalues $-d$ and $d$, respectively.
Here, $d\equiv\sqrt{d_0^2+d_1^2}$ is related to the system's characteristic frequency $\omega_0 = 2d$.
The gauge choice for the eigenvectors is immaterial because the geometric phase factors are gauge invariant.

For a pure state $|\psi(t)\rangle$, the density operator $\rho(t)=|\psi(t)\rangle\langle\psi(t)|$ admits the Bloch representation
\begin{equation}
\rho(t)=\frac{1}{2}\Bigl(\Id+\boldsymbol{r}(t)\cdot\boldsymbol{\sigma}\Bigr),
\end{equation}
and the Bloch vector is given by
\begin{equation}
r_i(t)=\mathrm{Tr}\!\left[\rho(t)\sigma_i\right]=\langle\psi(t)|\sigma_i|\psi(t)\rangle.
\label{eq:bloch_def}
\end{equation}
The instantaneous energy expectation is, then, written as
\begin{equation}
\langle H(t)\rangle= -\boldsymbol{d}(t)\cdot\boldsymbol{r}(t).
\label{eq:energy_dot}
\end{equation}

\subsection{Geometric Phases}

For the finite-speed evolution generated by $H(t)$, the state need not return to the same path after one driving period. We therefore use the kinematic geometric phase for a noncyclic evolution with nonorthogonal endpoints~\cite{SamuelBhandari1988,MukundaSimon1993}. Let
\begin{equation}
|\psi(t)\rangle=U(t)|\psi_0\rangle,
\qquad
|\psi_0\rangle=|g(0)\rangle .
\label{eq:phi_tot_def}
\end{equation}
Whenever $\langle\psi_0|\psi(T)\rangle\neq0$, the overlap defines the total phase
\begin{equation}
\phi_{\mathrm{tot}}\equiv\Argop\langle\psi_0|U(T)|\psi_0\rangle .
\end{equation}
Using Eqs.~\eqref{eq:U_exact}--\eqref{eq:UT} and the initial Bloch vector
$\boldsymbol{n}_0\equiv\langle\psi_0|\boldsymbol{\sigma}|\psi_0\rangle$, we obtain
\begin{equation}
\langle\psi_0|\psi(T)\rangle
=-\left[\cos(\OmR T)+i\sin(\OmR T)\frac{\boldsymbol{v}\cdot\boldsymbol{n}_0}{\OmR}\right].
\label{eq:overlap_exact}
\end{equation}
The dynamical phase is
\begin{equation}
\phi_{\mathrm{dyn}}
\equiv-\int_0^T\exd t\,\langle\psi(t)|H(t)|\psi(t)\rangle
=\int_0^T\exd t\,\boldsymbol{d}(t)\cdot\boldsymbol{r}(t).
\label{eq:phi_dyn_def}
\end{equation}
Let $\Rotz(\omega t)$ denote the $SO(3)$ rotation associated with the $SU(2)$ unitary $R(t)$ of Appendix~\ref{App1}, where $R(t)=\exp(-i\omega t\sigma_z/2)$. Then
$\boldsymbol{d}(t)=\Rotz(\omega t)\boldsymbol{d}_{\mathrm{static}}$ and
$\boldsymbol{r}(t)=\Rotz(\omega t)\boldsymbol{r}_{\mathrm{rot}}(t)$, so
\begin{equation}
\boldsymbol{d}(t)\cdot\boldsymbol{r}(t)
=\boldsymbol{d}_{\mathrm{static}}\cdot\boldsymbol{r}_{\mathrm{rot}}(t)
\label{eq:dot_invariant}
\end{equation}
by $\Rotz^{\mathsf T}\Rotz=\Id$. For the initial ground state,
\begin{equation}
\boldsymbol{n}_0=\frac{(d_0,0,d_1)}{d}
=\frac{\boldsymbol{d}_{\mathrm{static}}}{d},
\qquad
\rho(0)=\frac12\bigl(\Id+\boldsymbol{n}_0\cdot\boldsymbol{\sigma}\bigr),
\label{eq:n0_ground}
\end{equation}
so $\boldsymbol{d}_{\mathrm{static}}=d\,\boldsymbol{n}_0$.

\subsection{Berry phase}

The kinematic geometric phase is
\begin{equation}
\gamma_{\mathrm g}\equiv\phi_{\mathrm{tot}}-\phi_{\mathrm{dyn}}.
\label{eq:gamma_geom}
\end{equation}
In the adiabatic limit $\eta\equiv\omega/\omega_0\ll1$, the evolved ray follows the instantaneous ground-state loop and $\gamma_{\mathrm g}$ approaches its Berry phase. With
\begin{equation}
\OmR=d+\frac{d_1}{d}\frac{\omega}{2}+\mathcal O(\omega^2/d),
\qquad
c\equiv\frac{\boldsymbol{v}\cdot\boldsymbol{n}_0}{\OmR}=1+\mathcal O(\eta^2),
\end{equation}
one has $\phi_{\mathrm{dyn}}=dT+\mathcal O(\eta)$. Choosing the branch of the total phase that is continuous in the adiabatic limit,
\begin{equation}
\phi_{\mathrm{tot}}
=\OmR T-\pi
=dT+\pi\left(\frac{d_1}{d}-1\right)+\mathcal O(\eta)
\quad(\mathrm{mod}\ 2\pi).
\end{equation}
Therefore,
\begin{equation}
\gamma_{\mathrm g}\longrightarrow\gamma_B
=\pi\left(\frac{d_1}{d}-1\right)
=\pi(\cos\theta-1)
=-\frac12\Omega[C]
\quad(\mathrm{mod}\ 2\pi),
\label{eq:berryphase_conical}
\end{equation}
where $C$ is the closed path traced by $\hat{\boldsymbol{n}}(t)=\boldsymbol{d}(t)/d$. We use $\Omega[C]$ for signed solid angles of Bloch-sphere curves throughout.

More generally, for a smooth family of normalized eigenstates $|n(\boldsymbol{d})\rangle$ over the control manifold, the Berry connection is the real-valued one-form
\begin{equation}
A_B \equiv i\,\langle n|\,\exd\,|n\rangle ;
\label{eq:berry-connection}
\end{equation}
the connection itself is gauge dependent, but its curvature two-form
\begin{equation}
F_B=\exd A_B
\label{eq:berry-curvature}
\end{equation}
and the holonomy $\gamma_B=\oint_C A_B$ of a closed loop are gauge invariant~\cite{Berry1984,Simon1983}. The adiabatic limit of Eq.~\eqref{eq:gamma_geom} is precisely this holonomy, with the loop supplied by $\boldsymbol{d}(t)$. For the two-level family of Eq.~\eqref{eq:H_lab} the space of field directions is the unit sphere, and in the gauge of Eq.~\eqref{eq:eigenkets_singlephase} the ground-state connection takes the monopole form
\begin{equation}
A_B=-\tfrac12\,(1-\cos\theta)\,\exd\phi,
\qquad
F_B=-\tfrac12\,\sin\theta\,\exd\theta\wedge\exd\phi=-\tfrac12\,\exd\Omega,
\label{eq:berry-monopole}
\end{equation}
the field of a magnetic monopole of charge $-1/2$ located at the degeneracy $\boldsymbol{d}=0$~\cite{Dirac1931}. Stokes' theorem then gives $\gamma_B=\oint_C A_B=-\frac12\Omega[C]$ for every closed loop $C$ on the sphere, of which Eq.~\eqref{eq:berryphase_conical} is the constant-latitude case.

\section{Uhlmann Holonomy}

\subsection{Uhlmann formalism}
\label{sec:uhlmann-formalism}

In this section, we summarize the Uhlmann holonomy, one definition of a geometric phase for mixed states~\cite{Uhlmann1986,Hubner1993,Viyuela2014_2D,Viyuela2015SPT,Viyuela2018Obs,Rivas2013}.
An amplitude matrix $w$ of a density operator $\rho$ is defined by
\begin{equation}
\rho \equiv w w^\dagger ,
\label{eq:uhlmann-amplitude}
\end{equation}
with the gauge freedom $w\mapsto wU$ for $U\in \mathbb U(n)$.

Along a smooth closed trajectory $\rho(\phi)$, parametrized by a generic loop parameter $\phi\in[0,2\pi]$ (realized below as $\phi=\omega t$ for the driven qubit), we work in terms of the square-root density operator and its gauge;
\begin{equation}
w(\phi)=\sqrt{\rho(\phi)}\,U(\phi),\qquad U(0)=\Id,
\label{eq:sqrt-gauge-U}
\end{equation}
and define the Uhlmann connection in the convention
\begin{equation}
A_U\equiv\exd U\,U^{-1}=\exd U\,U^\dagger,
\qquad
\frac{\exd U(\phi)}{\exd\phi}=A_U(\phi)\,U(\phi).
\label{eq:uhlmann-transport-U}
\end{equation}
The associated Uhlmann holonomy around the loop is
\begin{equation}
U_C\equiv \mathcal{P}\exp\!\left(\oint A_U\right) ,
\label{eq:uhlmann-holonomy-UC}
\end{equation}
where $\mathcal{P}$ denotes the path ordering.

For full-rank trajectories, the Uhlmann parallel-transport condition fixes the connection,
\begin{equation}
\{\rho,A_U\}=\bigl[\exd\sqrt{\rho},\sqrt{\rho}\bigr],
\label{eq:uhlmann-parallel}
\end{equation}
where $\{\cdot,\cdot\}$ is the anticommutator, and $[\cdot,\cdot]$ is the commutator.

The corresponding Uhlmann curvature as a two-form is defined by
\begin{equation}
F_U=\exd A_U-A_U\wedge A_U,
\label{eq:uhlmann-curvature}
\end{equation}
which captures the non-Abelian structure of the connection in the transport convention of Eq.~\eqref{eq:uhlmann-transport-U}. If the Uhlmann connection is Abelian, the quadratic term $A_U\wedge A_U$ vanishes. 
Consequently, the mathematical structures of the Uhlmann curvature and connection reduce to the same form as those of the Berry curvature and connection.

A gauge-invariant Uhlmann geometric phase is given by
\begin{equation}
\gamma_U := \Argop~\mathrm{Tr}\!\left[w(0)^\dagger w(2\pi)\right].
\label{eq:uhlmann-gamma-def}
\end{equation}
With Eq.~\eqref{eq:sqrt-gauge-U}, this becomes
\begin{equation}
\gamma_U=\Argop~\mathrm{Tr}\!\left[\sqrt{\rho(0)}\,U_C\,\sqrt{\rho(2\pi)}\right],
\label{eq:uhlmann-gamma-sqrt}
\end{equation}
For a closed trajectory, $\rho(2\pi)=\rho(0)$, and therefore
\begin{equation}
\gamma_U=\Argop~\mathrm{Tr}\!\left[\rho(0)\,U_C\right].
\label{eq:uhlmann-gamma-closed}
\end{equation}

\subsection{Uhlmann Phase of The Driven Qubit System}
Assuming that the system has reached thermal equilibrium through its interaction with the heat bath at $T_B$, the Gibbs state of the driven qubit system is
\begin{equation}
\rho(\phi)=\frac{e^{-\beta H(\phi)}}{\mathrm{Tr}\,e^{-\beta H(\phi)}}
=\frac{1}{2}\Big(\Id+r\,\hat{\boldsymbol{n}}(\phi)\cdot\boldsymbol{\sigma}\Big),
\label{eq:rho_Gibbs_form}
\end{equation}
with Bloch radius
\begin{align}
r=\tanh\!\big(\beta d\big),
\label{eq:BlochVec_Gibbs}
\end{align}
where the inverse temperature $\beta \equiv 1/ T_B$, and the unit Bloch vector
\begin{equation}
\hat{\boldsymbol{n}}(\phi)=\frac{\boldsymbol{d}(\phi)}{d}
=(\sin\theta\cos\phi,\ \sin\theta\sin\phi,\ \cos\theta).
\label{eq:nhat_loop}
\end{equation}
Here the loop parameter is realized physically as $\phi=\omega t$.
Because the scalar magnitude $d$ is constant on the loop, $r$ is constant as well. The eigenvalues and eigenprojectors of $\rho(\phi)$ are
\begin{equation}
\lambda_\pm=\frac{1\pm r}{2},\qquad
P_\pm(\phi)=\frac{1}{2}\Big(\Id\pm\hat{\boldsymbol{n}}(\phi)\cdot\boldsymbol{\sigma}\Big).
\label{eq:eigs_proj}
\end{equation}
and the eigenprojectors satisfy completeness, $P_+ + P_-=\Id$.

Using the amplitude convention and the anti-Hermiticity of the Uhlmann
connection, we solve Eq.~\eqref{eq:uhlmann-parallel} algebraically to obtain
(see Appendix~\ref{App:UhlCon} for details)
\begin{equation}
A_U
=-\frac{i}{2}\,f(r)\,(\hat{\boldsymbol{n}}\times \exd\hat{\boldsymbol{n}})\cdot\boldsymbol{\sigma}.
\label{eq:AU_final_copy}
\end{equation}
For the conical drive traversed over one period $T$, $\theta$ is constant and $\exd\hat{\boldsymbol{n}}=(\partial_\phi\hat{\boldsymbol{n}})\,\exd\phi$ with
\begin{equation}
\hat{\boldsymbol{n}}\times\partial_\phi\hat{\boldsymbol{n}}
=\big(-\cos\theta\sin\theta\cos\phi,\ -\cos\theta\sin\theta\sin\phi,\ \sin^2\theta\big).
\label{eq:cross_copy}
\end{equation}
Therefore, the one-form vector of the Uhlmann connection $A_U$ is written as $A_U(\phi)\,\exd\phi$ where
\begin{equation}
A_U(\phi)
=-\frac{i}{2}\,f(r)\Big[
-\cos\theta\sin\theta(\cos\phi\,\sigma_x+\sin\phi\,\sigma_y)
+\sin^2\theta\,\sigma_z
\Big].
\label{eq:Aphi_copy}
\end{equation}
The transport operator in Eq.~\eqref{eq:uhlmann-holonomy-UC} is generated by the path-ordered exponential
\begin{equation}
U(\phi) = \mathcal{P} \exp\left( \int_0^\phi A_U(\phi') \,\exd\phi' \right)
\label{eq:path-order}
\end{equation}
The explicit form of this operator is derived in Appendix~\ref{App:UhlHol}. 
Thus, the Uhlmann holonomy accumulated over a single driving period, $U_C = U[\phi(T)=2\pi]$, takes the form
\begin{equation}
U_C = - \left[ \cos\left(\pi\Theta\right) \Id + i \frac{\sin\left(\pi\Theta\right)}{\Theta} \Big(f(r)\cos\theta\sin\theta \sigma_x + (1-f(r)\sin^2\theta)\sigma_z \Big) \right],
\label{eq:UC_final_copy}
\end{equation}
where $\Theta$ denotes the norm of the effective field vector governing the holonomy,
\begin{equation}
\Theta = \sqrt{\sin^2\theta\left(f(r)^2 - 2f(r)\right) + 1}.
\label{eq:Theta_main_copy}
\end{equation}
For a closed loop, the gauge-invariant complex scalar is
\begin{align}
\mathcal{G}
&\equiv
\mathrm{Tr}\!\left[\sqrt{\rho(0)}\,U_C\,\sqrt{\rho(2\pi)}\right]\notag\\
&=-\left[\cos\left(\pi\Theta\right)
+i r\cos\theta\frac{\sin\left(\pi\Theta\right)}{\Theta}\right].
\label{eq:G_gibbs}
\end{align}

\begin{figure}
    \centering
    \includegraphics[width=1\linewidth]{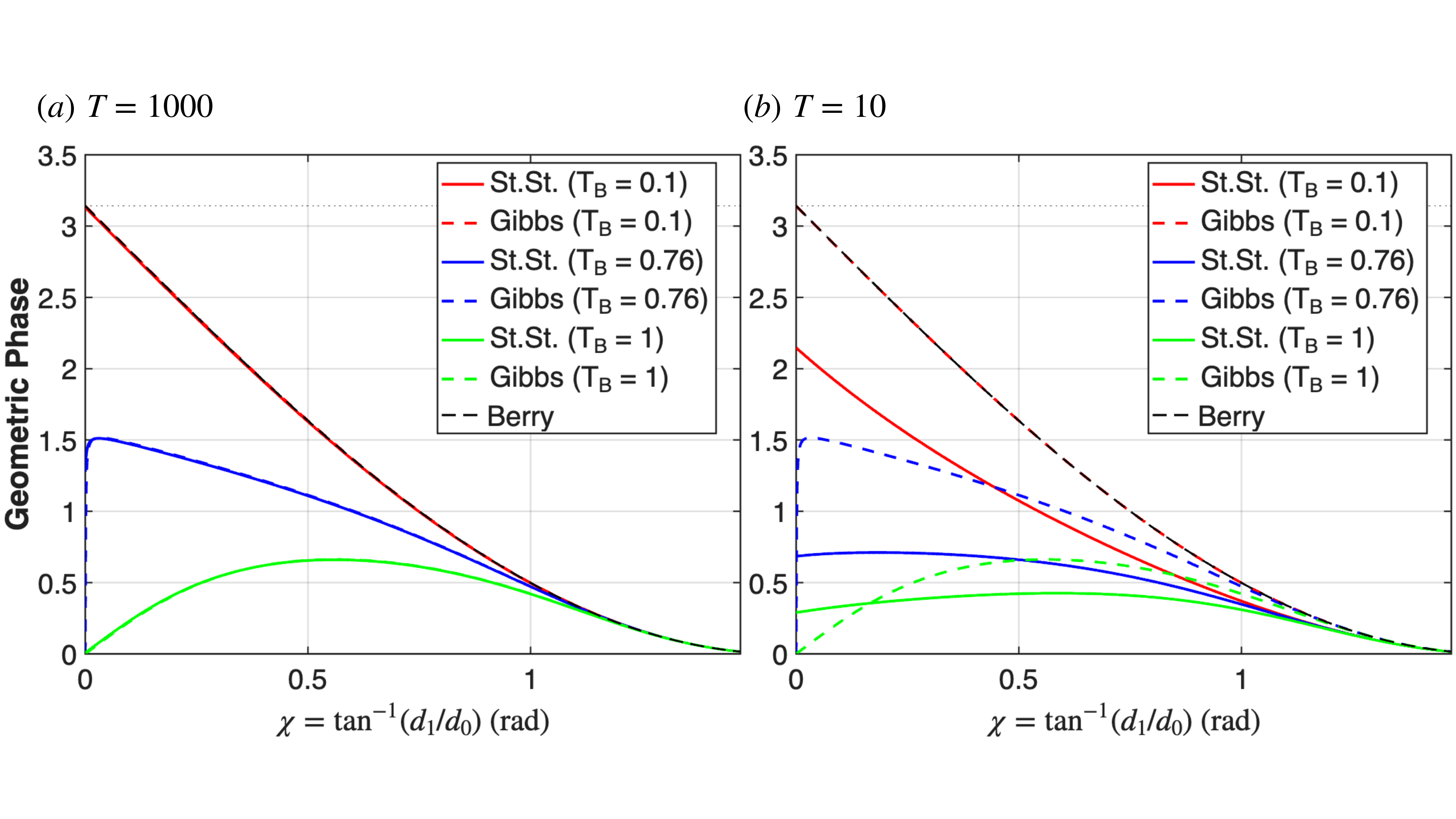}
    \caption{Magnitudes of geometric phases in a physical $d_1$ scan at fixed $d_0=1$, with $\chi=\tan^{-1}(d_1/d_0)$. The equatorial loop is $\chi=0$ and the polar limit is $\chi\to\pi/2$. Dashed curves: equilibrium Gibbs loop; solid curves: dynamical steady-state limit cycle, for $T_B=0.1,\,0.76,\,1.0$; black dashed curve: Berry phase $\pi(1-\sin\chi)$. (a)~Near-adiabatic driving, $T=1000$: solid and dashed curves coincide, $\gamma_U^{\rm dyn}\to\gamma_U^{\rm eq}$. (b)~Fast driving, $T=10$: the lag separates the two. Near $T_C\simeq0.759$, the equatorial value exhibits the discontinuity of the Uhlmann phase.}
    \label{fig:Uhlmann_Phase}
\end{figure}

Therefore, the Uhlmann phase is defined by 
\begin{equation}
\gamma_U \equiv {\Argop}~\mathcal{G}.
\label{eq:Phi_final_copy}
\end{equation}
Evaluating this expression as a function of the field angle $\theta$ and the bath temperature yields the dashed curves in Fig.~\ref{fig:Uhlmann_Phase}. Its abscissa is $\chi=\tan^{-1}(d_1/d_0)=\pi/2-\theta$. We fix $d_0=1$ and vary $d_1$. At low $T_B$ the Uhlmann phase tracks the Berry curve over most of the range, while at high $T_B$ it is progressively suppressed and goes to zero.

On the equatorial path $(\theta=\pi/2)$,
Eq.~\eqref{eq:Aphi_copy} reduces to
\begin{equation}
A_U=-\frac{i}{2}f(r)\sigma_z\,\exd\phi,
\label{eq:AU_equator_copy}
\end{equation}
and Eq.~\eqref{eq:UC_final_copy} gives
\begin{equation}
U_C
=\exp\!\big(-i\pi f(r)\sigma_z\big).
\label{eq:UC_equator_copy}
\end{equation}
This result shows that the Uhlmann holonomy loses its dependence on $\sigma_x$ and $\sigma_y$, and thus reduces to an effective Abelian form that is insensitive to path ordering.
Finally, Eq.~\eqref{eq:Phi_final_copy} yields
\begin{equation}
\gamma_U={\Argop}\Big[\cos\!\big(\pi f(r)\big)\Big].
\label{eq:Phi_equator_copy}
\end{equation}
The phase is restricted to the two values $0~(T_B > T_C)$ and $\pi~(T_B < T_C)$, separated by a critical temperature $T_C \simeq 0.759$ at which $\cos(\pi f(r))=0$. Since the equatorial holonomy is effectively Abelian, this temperature-driven change is a discontinuous Uhlmann-phase jump, in contrast to the continuous behavior away from the equator ($\theta \neq \pi/2$), where the non-Abelian components survive. This discontinuity appears in Fig.~\ref{fig:Uhlmann_Phase} as the abrupt change of the low-temperature curves near the equatorial point $d_1\to0$.

Both constructions presuppose a strictly positive gap, which enters through the unit direction $\hat{\boldsymbol{n}}$ and the Gibbs radius $r=\tanh(\beta d)$; the gap-closing case is taken up in Sec.~\ref{sec:gapclosing}.

\section{Dynamical Approach}
\label{sec:dyn}

We now treat the qubit as an open system and solve the driven Lindblad dynamics in closed form. This yields (i) the exact rotating-frame propagator, (ii) the nonequilibrium steady state, which corresponds to a limit cycle replacing the instantaneous Gibbs loop, and (iii) the limits in which the equilibrium description of Sec. III is recovered, with the associated geometric phase reducing to the Berry phase at zero temperature.

\subsection{Optical master equation and Bloch representation}
\label{sec:ome}

We take a system--bath Hamiltonian $H_{\rm tot}(t)=H(t)\otimes\Id_B+\Id_S\otimes H_B+H_{SB}$ and
trace out the bath. Under the Born--Markov and secular approximations the reduced density
operator obeys the optical master equation
\begin{equation}
\frac{\exd\rho}{\exd t}=-i[H(t),\rho]
+\!\!\sum_{j\in\{\downarrow,\uparrow\}}\!\!\gamma_j\,\mathcal D[L_j]\rho
+\frac{\gamma_\phi}{2}\,\mathcal D[L_z]\rho,
\label{eq:gksl_general}
\end{equation}
with $\mathcal D[L]\rho=L\rho L^\dagger-\tfrac12\{L^\dagger L,\rho\}$. Energy relaxation is
implemented with instantaneous ladder operators referred to the momentary eigenbasis,
\begin{equation}
L_\downarrow(t)=|g(t)\rangle\langle e(t)|,\quad
L_\uparrow(t)=|e(t)\rangle\langle g(t)|,
\label{eq:ladder}
\end{equation}
and we allow an optional pure-dephasing channel
$L_z(t)=|e(t)\rangle\langle e(t)|-|g(t)\rangle\langle g(t)|$ (the original ladder-only model is
recovered at $\gamma_\phi=0$). 

For a bosonic bath at temperature $T_B$, let $\nth=(e^{\beta\omega_0}-1)^{-1}$ and $\gamma\equiv 2\pi J(\omega_0)$, where $J(\omega)$ denotes the Drude--Lorentz spectral density
$J(\omega)=2\lambda\gamma_D\omega/(\omega^2+\gamma_D^2)$. Here, $\lambda$ and $\gamma_D$ are the coupling strength and cutoff frequency, respectively. The rates are then
\begin{equation}
\gamma_\downarrow=\gamma(\nth+1),\qquad
\gamma_\uparrow=\gamma\,\nth.
\label{eq:rates}
\end{equation}

Writing $\rho=\tfrac12(\Id+\boldsymbol{r}\cdot\boldsymbol{\sigma})$, the coherent part of the evolution gives the precession
$\dot{\boldsymbol{r}}|_{\rm coh}=2\,\boldsymbol{r}\times\boldsymbol{d}(t)$. The field-adapted basis is the right-handed orthonormal triad
$\hat{\boldsymbol{e}}_3(t)=\hat{\boldsymbol{n}}(t)$,
$\hat{\boldsymbol{e}}_2(t)=[\hat{\boldsymbol{z}}\times\hat{\boldsymbol{n}}(t)]/\sin\theta$, and
$\hat{\boldsymbol{e}}_1(t)=\hat{\boldsymbol{e}}_2(t)\times\hat{\boldsymbol{e}}_3(t)$. In this co-rotating basis, the dissipators produce relaxation toward the instantaneous Gibbs state with a diagonal rate tensor (see Appendix~E),
\begin{equation}
\Geff=\mathrm{diag}\big(1/T_2,\,1/T_2,\,1/T_1\big),
\label{eq:relax}
\end{equation}
with $1/T_1=\gamma(2\nth+1)=\gamma_\downarrow+\gamma_\uparrow$ and $1/T_2=1/(2T_1)+\gamma_\phi$.
The longitudinal (field-axis) component carries the population difference; the transverse
coherences decay at half the longitudinal rate plus pure dephasing, so that $T_2\le2T_1$.
Collecting both contributions, the
laboratory-frame Bloch equation is
\begin{equation}
\dot{\boldsymbol{r}}(t)=2\,\boldsymbol{r}\times\boldsymbol{d}(t)-\Geff_{\rm lab}(t)\big(\boldsymbol{r}-\boldsymbol{r}_{\rm eq}(t)\big),\quad
\boldsymbol{r}_{\rm eq}(t)=\req\,\hat{\boldsymbol{n}}(t),
\label{eq:blocheq}
\end{equation}
with $\req=\tanh(\beta d)$ and $\Geff_{\rm lab}(t)=\mathcal R(t)\,\Geff\,\mathcal R(t)^{\mathsf T}$.
The rotation $\mathcal R(t)$, whose columns are the field-adapted axes $\hat{\boldsymbol{e}}_i(t)$, carries the
field-adapted frame into the laboratory frame, so $\Geff_{\rm lab}(t)$ is
explicitly time dependent; the generator is therefore not autonomous in the laboratory
frame. This time dependence is removed in the rotating frame.

\subsection{Rotating-frame generator and exact propagator}
\label{sec:rot}

Following Appendix~A, the rotation $\Rotz(\omega t)$ renders the field static. In the rotating
frame the Bloch vector $\boldsymbol{r}_{\rm rot}=\Rotz(-\omega t)\boldsymbol{r}$ obeys the autonomous linear
equation
\begin{equation}
\dot{\boldsymbol{r}}_{\rm rot}=M\,\boldsymbol{r}_{\rm rot}+\Geff\,\boldsymbol{r}_{\rm eq},\qquad
M=[\boldsymbol{\Omega}]_\times-\Geff,
\label{eq:autonomous}
\end{equation}
with precession vector $\boldsymbol{\Omega}=-2\boldsymbol{v}$ of Eq.~\eqref{eq:vOmega}, where $\boldsymbol{v}=(d_0,0,\Delta)$ and
$\Delta=d_1+\tfrac\omega2$, and $\boldsymbol{r}_{\rm eq}=\req\hat{\boldsymbol{n}}_0$ now stationary.
In the field-adapted basis $\boldsymbol{\Omega}=(\Omega_1,0,\Omega_3)$, with
\begin{equation}
\Omega_1=\omega\sin\theta,\quad
\Omega_3=-\big(\omega_0+\omega\cos\theta\big),
\label{eq:omegacomp}
\end{equation}
where $\OmR=|\boldsymbol{v}|$ is the effective Rabi frequency and the Bloch vector precesses at the level splitting $|\boldsymbol{\Omega}|=2\OmR$. The generator is therefore
\begin{equation}
M=\begin{pmatrix}-a&-\Omega_3&0\\ \Omega_3&-a&-\Omega_1\\ 0&\Omega_1&-b\end{pmatrix},
\qquad a\equiv\frac1{T_2},\ b\equiv\frac1{T_1}.
\label{eq:Mmatrix}
\end{equation}

The eigenvalues of $M$ govern the relaxation generated by the propagator $e^{Mt}$. Except at isolated parameter values where two eigenvalues coalesce, they are either all real or consist of one real eigenvalue and one complex-conjugate pair. The former case produces nonoscillatory relaxation, whereas the latter produces a decay along the real eigendirection together with damped oscillations in the complementary plane. The boundary between the two regimes is exactly this coalescence locus, the vanishing of the discriminant of the characteristic polynomial. A sufficient condition for the underdamped regime is $\sqrt3\,|\boldsymbol{\Omega}|>|1/T_2-1/T_1|$ (see Appendix~F), which in the weak system-bath coupling regime relevant here is satisfied by a wide margin: $\gamma\ll\omega_0$ gives $|\boldsymbol{\Omega}|\sim\omega_0$ against $|1/T_2-1/T_1|\sim\gamma$. The spectrum therefore consists of one real eigenvalue and one complex-conjugate pair,
\begin{equation}
\mu_1=-\Gamma_L,\qquad \mu_{2,3}=-\Gamma_T\pm i\,\Omega_{\rm eff}.
\label{eq:eig}
\end{equation}
Their closed (Cardano) forms are given in Appendix~F; 
to leading order in
$\gamma/\omega_0$ and in the near-adiabatic limit $\eta\equiv\omega/\omega_0\ll1$, with
corrections entering only at $O(\eta^{2})$,
\begin{equation}
\Gamma_L\simeq\frac1{T_1},\quad
\Gamma_T\simeq\frac1{T_2},\quad
\Omega_{\rm eff}\simeq\omega_0+\omega\cos\theta,
\label{eq:eigadiab}
\end{equation}
so the three eigenvalues become the physical relaxation rates and the dressed precession
frequency. The exact propagator follows from the Lagrange--Sylvester expansion
$e^{Mt}=\sum_k e^{\mu_k t}P_k$ and can be written in real form (see Appendix~F) as a
longitudinal decay along $P_L$ plus a damped rotation in the complementary plane. The general
solution of Eq.~\eqref{eq:autonomous} is
\begin{equation}
\boldsymbol{r}_{\rm rot}(t)=\boldsymbol{r}_{\rm ss}+e^{Mt}\big(\boldsymbol{r}_{\rm rot}(0)-\boldsymbol{r}_{\rm ss}\big),
\label{eq:transient}
\end{equation}
where $\boldsymbol{r}_{\rm ss}=-M^{-1}\Geff\,\boldsymbol{r}_{\rm eq}$ is the unique fixed point.

Transforming back, $\boldsymbol{r}_{\rm lab}(t)=\Rotz(\omega t)\boldsymbol{r}_{\rm rot}(t)$. After the transient decays, the qubit settles onto a constant-latitude, constant-radius limit cycle.

\subsection{Steady-state limit cycle}
\label{sec:ss}

\begin{figure}
    \centering
    \includegraphics[width=1\linewidth]{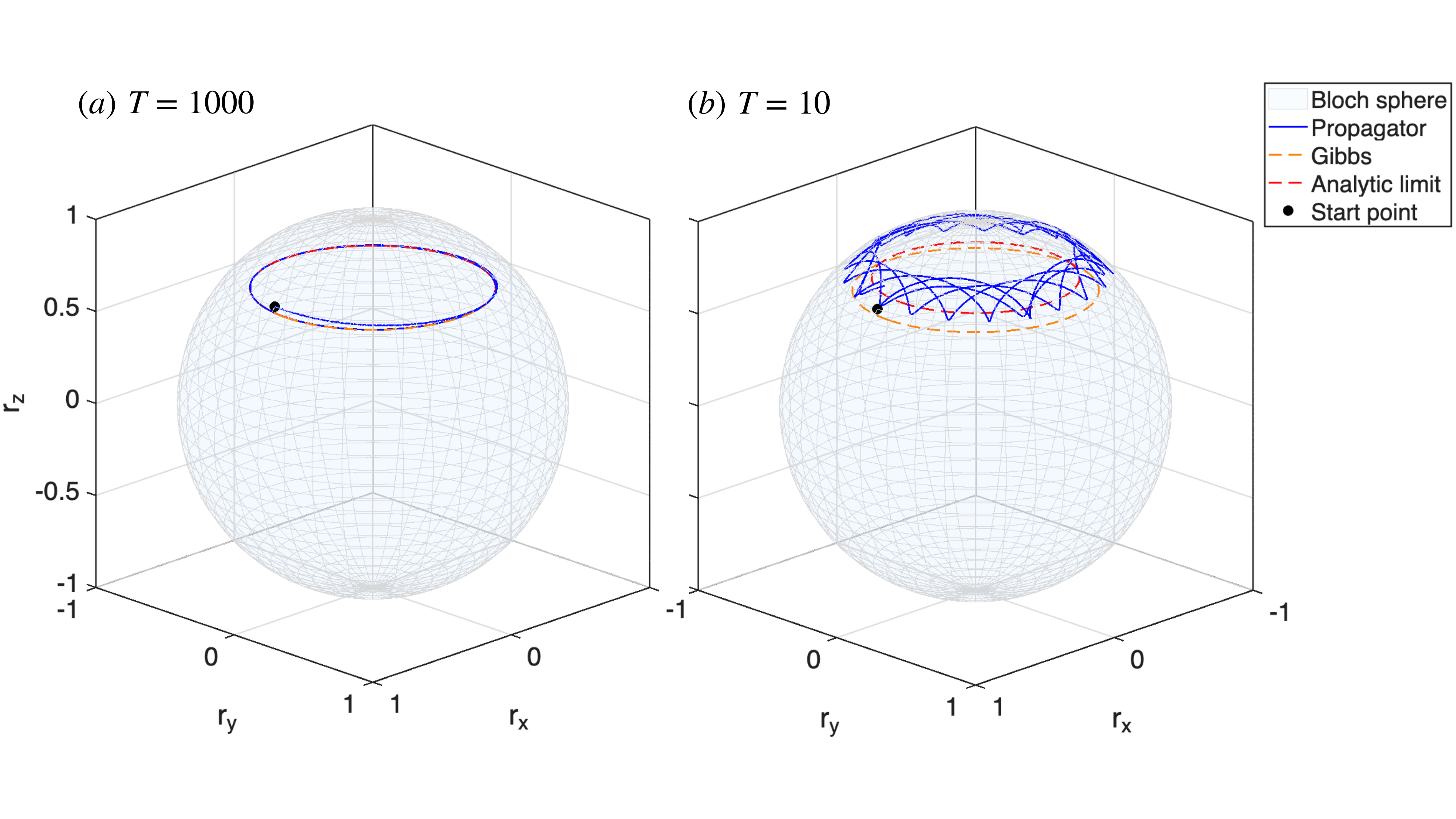}
    \caption{Bloch-sphere trajectories of the driven qubit, computed from the exact rotating-frame solution, transformed to the laboratory frame, for $d_0=d_1=1$ and $T_B=1$, with Drude spectral density of coupling strength $\lambda=0.001$ and cutoff frequency $\gamma_D=5$. Solid curve: the propagator trajectory from the initial state (dot); red dashed curve: the analytic limit cycle $\boldsymbol{r}_{\rm ss}$; orange dashed curve: the instantaneous Gibbs trajectory. (a)~Near-adiabatic driving, $T=1000$: after a short transient the three curves are indistinguishable, and the steady state tracks the Gibbs loop. (b)~Fast driving, $T=10$: the trajectory spirals onto a limit cycle that is visibly tilted and displaced from the Gibbs loop due to the lag.}
    \label{fig:Trajectory}
\end{figure}

For $T_1,T_2>0$, the autonomous generator has a unique steady state $\boldsymbol r_{\rm ss}$. 
For $\omega\neq0$, this state generally does not coincide with the instantaneous Gibbs state. Indeed, the rotating-frame contribution $-\omega\hat{\boldsymbol{z}}$ shifts the precession vector $\boldsymbol{\Omega}$ from $-2\boldsymbol{d}_{\rm static}$ to $-2\boldsymbol{d}_{\rm static}-\omega\hat{\boldsymbol{z}}$, so it is no longer parallel to the field axis $\hat{\boldsymbol{n}}_0$. This misalignment produces the residual torque $\boldsymbol{\Omega}\times\boldsymbol{r}_{\rm eq}
=-r_{\rm eq}\Omega_1\hat{\boldsymbol{e}}_2$, which must be balanced by dissipation in the steady state.
Writing $\boldsymbol{r}_{\rm ss}=\boldsymbol{r}_{\rm eq}+\boldsymbol{\delta}$ with $\boldsymbol{r}_{\rm eq}=\req\hat{\boldsymbol{e}}_3$ and solving the stationarity condition $\boldsymbol{\Omega}\times\boldsymbol{r}_{\rm ss}=\Geff\boldsymbol{\delta}$ component-wise, as detailed in Appendix~G,
\begin{equation}
\delta_1=\frac{\req\Omega_1\Omega_3T_2}{D},\quad
\delta_2=\frac{-\req\Omega_1}{D},\quad
\delta_3=\frac{-\req\Omega_1^2T_1}{D},
\label{eq:lag}
\end{equation}
with $D=1/T_2+\Omega_1^2T_1+\Omega_3^2T_2$. The limit-cycle geometry is fixed by the radius and polar angle
\begin{equation}
\begin{gathered}
\rss=\sqrt{\delta_1^2+\delta_2^2+(\req+\delta_3)^2},\\[2pt]
\cos\tss=\frac{(\req+\delta_3)\cos\theta-\delta_1\sin\theta}{\rss},
\end{gathered}
\label{eq:rssthetass}
\end{equation}
together with an azimuthal offset
$\phi_0=\operatorname{atan2}\!\big(\delta_2,\,\delta_1\cos\theta+(\req+\delta_3)\sin\theta\big)$.
A nonzero $\delta_2$ means the steady Bloch vector lags the instantaneous field in azimuth, which is the essential distinction between the dynamical limit cycle and the Gibbs cycle.

Expressed in control parameters (see Appendix~G), the three lags scale as positive powers of $\eta$,
\begin{equation}
\delta_1\simeq-\req\,\eta\sin\theta,\quad
\delta_2\simeq-\frac{\req\,\eta\sin\theta}{T_2\,\omega_0},\quad
\delta_3\simeq-\req\,\eta^2\sin^2\theta\,\frac{T_1}{T_2},
\label{eq:lagadiab}
\end{equation}
so that $\boldsymbol{r}_{\rm ss}\to\boldsymbol{r}_{\rm eq}$ as $\eta\to0$, for any fixed bath temperature. The
tilt $\delta_1=\mathcal O(\eta)$ is the dominant correction and sets the rate of Gibbs convergence; this is the quantitative reason the dissipative trajectory tracks the Gibbs trajectory in the slow-driving regime, as illustrated in Fig.~\ref{fig:Trajectory}. In this limit the open-system dynamics therefore
contains the equilibrium description as its fixed point: the instantaneous Gibbs loop assumed
in Sec.~III is recovered as the $\eta\to0$ limit of the dynamical limit cycle,
establishing the physical consistency of the two treatments.

\subsection{\texorpdfstring{Zero-temperature limit: effective $U(2)\to U(1)$ reduction and Berry correspondence}{Zero-temperature limit: U(2) to U(1) reduction and Berry correspondence}}
\label{sec:collapse}

The Uhlmann holonomy $U_C$ and phase $\gamma_U=\Argop\,\mathrm{Tr}[\rho(0)U_C]$ derived in Sec.~III,
Eqs.~\eqref{eq:UC_final_copy}--\eqref{eq:Phi_final_copy}, depend on the state only through its Bloch radius $r$ and its cone angle $\vartheta$, the
polar angle of the state axis:
\begin{equation}
\begin{gathered}
\gamma_U=\Argop\!\Big\{\!-\!\Big[\cos(\pi\Theta)+i\,r\cos\vartheta\,\tfrac{\sin(\pi\Theta)}{\Theta}\Big]\Big\},
\end{gathered}
\label{eq:gammaU}
\end{equation}
with $\Theta=\sqrt{1-r^2\sin^2\vartheta}$.
The equilibrium (Gibbs) loop in the adiabatic limit uses $(r,\vartheta)=(\req,\theta)$ and reproduces
Eqs.~\eqref{eq:G_gibbs}--\eqref{eq:Phi_final_copy}. The dynamical limit cycle is again a circle of constant radius and constant latitude traversed at a uniform rate. The constant azimuthal offset $\phi_0$ merely shifts the starting point of the loop and leaves the Uhlmann phase unchanged. The derivation of Sec.~III applies verbatim with $(r,\vartheta)=(\rss,\tss)$,
\begin{equation}
\gamma_U^{\rm dyn}=\Argop\!\Big\{\!-\!\Big[\cos(\pi\Theta_{\rm ss})+i\,\rss\cos\tss\,
\tfrac{\sin(\pi\Theta_{\rm ss})}{\Theta_{\rm ss}}\Big]\Big\},
\label{eq:gammaUdyn}
\end{equation}
with $\Theta_{\rm ss}=\sqrt{1-\rss^2\sin^2\tss}$. The two expressions have the same form but differ through the lags in Eq.~\eqref{eq:lagadiab}: $\rss=\req+\mathcal O(\eta^2)$ and $\tss=\theta+\mathcal O(\eta)$. 
Hence for $\eta,\gamma/\omega_0\ll1$ the dynamical phase reduces to the equilibrium one, $\gamma_U^{\rm dyn}\to\gamma_U^{\rm eq}$; away from that regime the lag makes them differ. Figure \ref{fig:Uhlmann_Phase} displays this comparison explicitly. The dashed curves show the equilibrium Gibbs phases, whereas the solid curves show the dynamical steady-state phases. For the solid curves, the relaxation rate is evaluated as
$\gamma=2\pi J(\omega_0)$ with
$\lambda=0.001$ and $\gamma_D=5$, while the pure-dephasing rate is taken as $\gamma_\phi=4\pi T_BJ'(0)$.

For the dynamical limit cycle, an exact reduction to the Berry holonomy is most transparently obtained along the ordered route in which the cycle first approaches the Gibbs loop as $\eta\to0$, after which the resulting equilibrium loop is cooled as $T_B\to0$. At fixed nonzero $\eta$, a residual $\mathcal O(\eta^4)$ mixedness remains even at zero temperature. On the rank-one support reached in the ordered limit, the physically relevant Uhlmann transport reduces effectively from $U(2)$ to the Berry $U(1)$ holonomy. This reduction can be understood through three related features: the approach to purity, the alignment of the holonomy axis
with the state axis, and the diagonalization of the holonomy in the corresponding eigenbasis.

\emph{(i) Purity.}
The eigenvalues of $\rho_{\rm ss}$ are $(1\pm r_{\rm ss})/2$. Equations~\eqref{eq:lag} and \eqref{eq:rssthetass} give
\begin{equation}
r_{\rm ss}^{2}
=
r_{\rm eq}^{2}
\frac{K\left(K+\Omega_1^{2}T_2\right)}
     {\left(K+\Omega_1^{2}T_1\right)^2},
\qquad
K=\frac{1}{T_2}+\Omega_3^{2}T_2 .
\label{eq:rss-purity}
\end{equation}
In the adiabatic limit, $\Omega_1\to0$, and hence
$r_{\rm ss}\to r_{\rm eq}$. Subsequent cooling gives
$r_{\rm eq}=\tanh(\beta d)\to1$, so that the steady state becomes
pure in the ordered limit. It is also instructive to reverse the
order temporarily: at fixed nonzero $\eta$, the zero-temperature
conditions $r_{\rm eq}=1$, $\gamma_\phi=0$, and $T_2=2T_1$ give
\begin{equation}
1-r_{\rm ss}^{2}
=
\frac{\Omega_1^{4}T_1^{2}}{D^{2}}
\simeq
\frac14\eta^{4}\sin^{4}\theta.
\label{eq:purity}
\end{equation}
Thus, cooling alone leaves an $O(\eta^4)$ purity deficit, which
disappears when the adiabatic limit is also taken.
Accordingly, in the ordered limit, $\rho_{\rm ss}(t)$ approaches the rank-one projector associated with its own axis $\hat{\bs u}_{\rm ss}=\bs r_{\rm ss}/r_{\rm ss}$. This axis differs from that of the bare ground state by the $\mathcal O(\eta)$ tilt in Eq.~\eqref{eq:lag} and approaches it as $\eta\to0$.

The convergence can be made explicit entry by entry. On the limit cycle,
\begin{equation}
\rho_{\rm ss}(t)=\frac12
\begin{pmatrix}
1+\rss\cos\tss &
\rss\sin\tss\,e^{-i(\omega t+\phi_0)}
\\[2pt]
\rss\sin\tss\,e^{+i(\omega t+\phi_0)} &
1-\rss\cos\tss
\end{pmatrix}.
\label{eq:rhossmat}
\end{equation}
With the half-angle identities, the ground-state eigenket of Eq.~\eqref{eq:eigenkets_singlephase} gives the projector $|g(t)\rangle\langle g(t)|
=\tfrac12\bigl(\Id+\hat{\boldsymbol n}(t)\cdot\boldsymbol{\sigma}\bigr)$, whose diagonal and off-diagonal entries are $(1\pm\cos\theta)/2$ and
$(\sin\theta/2)e^{\mp i\omega t}$, respectively. Every entry of Eq.~\eqref{eq:rhossmat} therefore converges to its pure-state counterpart in the joint adiabatic and zero-temperature limit: as $\eta\to0$, one has $\rss\to r_{\rm eq}$, $\tss\to\theta$, and $\phi_0\to0$, while cooling gives $r_{\rm eq}\to1$. More explicitly, $\tss=\theta-\eta\sin\theta+O(\eta^2)$ and
$\phi_0=-\eta/(T_2\omega_0)+O(\eta^2)$. In particular, the coherences approach the finite pure-state values
$(\sin\theta/2)e^{\mp i\omega t}$ rather than vanishing. In this joint limit, $\rho(0)=|g(0)\rangle\langle g(0)|$, and hence the Uhlmann phase samples only the action of the holonomy on the one-dimensional occupied support,
\begin{equation}
\gamma_U=\Argop\langle g(0)|U_C|g(0)\rangle.
\end{equation}
The physically relevant transport therefore reduces effectively to $U(1)$.

\emph{(ii) Axis alignment.}
Evaluated on the limit cycle, Eq.~\eqref{eq:UC_final_copy} reads
\begin{equation}
U_C=-\Big[\cos(\pi\Theta)\,\Id+i\,\frac{\sin(\pi\Theta)}{\Theta}\,\boldsymbol m\cdot\boldsymbol\sigma\Big],
\quad
\boldsymbol m=\begin{pmatrix} f(r)\cos\vartheta\sin\vartheta\\ 0\\ 1-f(r)\sin^2\vartheta\end{pmatrix},
\label{eq:mvec}
\end{equation}
where $\boldsymbol m$ is the effective field vector governing the holonomy and $\Theta=|\boldsymbol m|$ reproduces Eq.~\eqref{eq:Theta_main_copy} with $\theta$ replaced by the state cone angle $\vartheta$.
In the pure-state limit $r\to1$, one has $f(r)=1-\sqrt{1-r^2}\to1$. Consequently, $\boldsymbol m \longrightarrow (\cos\vartheta\sin\vartheta,0,\cos^2\vartheta) = \cos\vartheta\,\hat{\boldsymbol n}$,
where
$\hat{\boldsymbol n}=(\sin\vartheta,0,\cos\vartheta)$ in the adapted coordinate system. Thus, $\boldsymbol m$ becomes collinear with the state axis and $\Theta=|\boldsymbol m|\to|\cos\vartheta|$. The holonomy accordingly reduces to
\begin{equation}
U_C\longrightarrow -\exp\left( i\pi\cos\vartheta\, \hat{\boldsymbol n}\cdot\boldsymbol{\sigma} \right),
\end{equation}
which is the Berry holonomy of a ground state traversing a cone of polar angle $\vartheta$.

\emph{(iii) Diagonalization.}
In the $\hat{\boldsymbol n}$ eigenbasis, $\hat{\boldsymbol n}\cdot\boldsymbol{\sigma} =\operatorname{diag}(1,-1)$, and hence
$U_C\longrightarrow \operatorname{diag} \left(-e^{i\pi\cos\vartheta}, -e^{-i\pi\cos\vartheta}\right)$.
Thus, the holonomy no longer mixes the occupied state with its
orthogonal complement. Projecting onto the occupied state with
$\rho(0)=|g\rangle\langle g|$ gives
\begin{equation}
\gamma_U
\longrightarrow
\Argop\left(-e^{i\pi\cos\vartheta}\right)
=
-\pi(1-\cos\vartheta)
\pmod{2\pi}.
\end{equation}
Only this occupied-state component contributes to the phase, so the
physically relevant holonomy reduces effectively to $U(1)$. Finally,
as $\eta\to0$, one has $\vartheta=\tss\to\theta$, and the resulting
Uhlmann phase coincides with the Berry phase $\gamma_B$ of the
instantaneous ground-state loop.

The dynamical Uhlmann phase therefore reduces to the Berry phase in the joint adiabatic and zero-temperature limit. Under the assumptions used here, the two limits commute, although taking the adiabatic limit first makes the intermediate equilibrium Gibbs loop explicit: $ \gamma_U^{\rm dyn} \xrightarrow{\eta\to0} \gamma_U^{\rm eq} \xrightarrow{T_B\to0} \gamma_B $. 
This correspondence, also illustrated in Fig.~\ref{fig:Uhlmann_Phase}, is restricted to uniformly gapped cycles. The weak-coupling/secular and adiabatic conditions $T_1^{-1},T_2^{-1}\ll\omega_0$ and $\omega\ll\omega_0$ must hold throughout the evolution. In addition, the exact autonomous rotating-frame solution used here requires the field magnitude $d$ to remain constant. If $d(t)$ vanishes at an isolated instant, the gap closes and these assumptions fail locally; the following section therefore compares the Berry and Uhlmann constructions through the crossing at the kinematic level.

\section{Geometric phases at a gap-closing degeneracy}
\label{sec:gapclosing}

\subsection{Transversal crossing of the degeneracy}
\label{sec:gap-setting}

Consider instead a smooth $T$-periodic control field $\boldsymbol{d}(t)$ whose magnitude $d(t)$ vanishes at a single instant $t_0$, with nonzero velocity $\boldsymbol{v}\equiv\dot{\boldsymbol{d}}(t_0)$---a transversal crossing. Writing $s\equiv t-t_0$ and $v=|\boldsymbol{v}|$, the eigenvalues $\mp d(t)$ meet linearly at the crossing. When plotted over any two-dimensional control plane through $\boldsymbol{d}=0$, the two energy surfaces form a double cone whose apex is the degeneracy---a \emph{diabolical point} in the terminology of Berry and Wilkinson~\cite{BerryWilkinson1984}. Because the instantaneous gap $\omega_0(t_0)$ vanishes, the secular
dynamical framework of Sec.~IV breaks down locally.
To leading order near $t_0$, $H(t)\simeq-s\,\boldsymbol v\cdot\boldsymbol{\sigma}$, which is the zero-gap limit of the Landau--Zener Hamiltonian~\cite{Landau1932,Zener1932}.
In the presence of the bath, the corresponding dynamics belongs to the dissipative Landau--Zener class~\cite{AoRammer1989,Wubs2006} and lies outside the
scope of this study. We therefore compare the two constructions at the kinematic level: the Berry phase of the instantaneous eigenstates and the Uhlmann phase of the instantaneous Gibbs family along the gap-closing drive.

\begin{figure}
    \centering
    \includegraphics[width=0.9\linewidth]{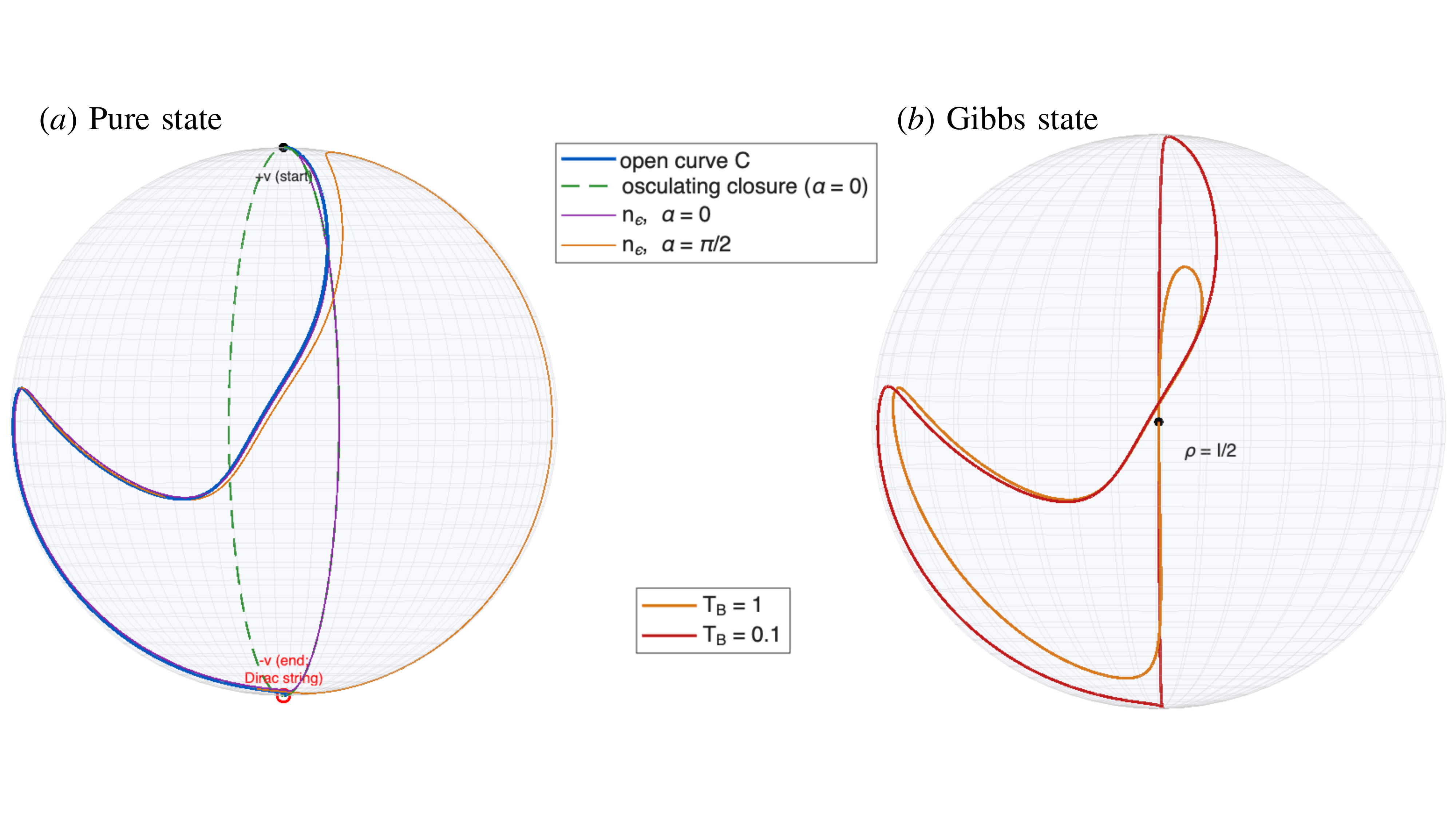}
    \caption{Bloch-space trajectories for the gap-closing drive of Eq.~\eqref{eq:demo-field}. (a)~On the unit sphere the direction $\hat{\boldsymbol{n}}=\boldsymbol{d}/d$ traces an open curve $C$ (blue) with antipodal endpoints $\pm\hat{\boldsymbol{v}}$. The dashed great circle is the osculating-plane closure ($\alpha=0$); the two thin curves are regularized loops $\hat{\boldsymbol{n}}_\epsilon$ with equal $\epsilon$ and closure azimuths $\alpha=0$ (purple) and $\alpha=\pi/2$ (yellow). (b)~The Gibbs trajectory $\boldsymbol{r}=\tanh(\beta d)\,\hat{\boldsymbol{n}}$, shown for $T_B=1$ (orange) and $0.1$ (red), is closed and smooth, passing through the maximally mixed state at the center of the Bloch sphere.}
    \label{fig:gap_paths}
\end{figure}

The pure- and mixed-state trajectories behave differently at the degeneracy. Near $t_0$ one has $\boldsymbol{d}(t)\simeq\boldsymbol{v}\,s$, so the unit direction obeys
\begin{equation}
\lim_{s\to0^\pm}\hat{\boldsymbol{n}}(t)=\pm\hat{\boldsymbol{v}},
\qquad
\hat{\boldsymbol{v}}\equiv\boldsymbol{v}/v .
\label{eq:njump}
\end{equation}
Removing $t_0$ from the periodic parameter cycle therefore produces an open curve $C$ on $S^2$, oriented from $+\hat{\boldsymbol v}$ just after the crossing to $-\hat{\boldsymbol v}$ just before the next passage through it. A Berry phase cannot be assigned to this curve without an additional closure prescription. The Gibbs-state trajectory, by contrast, remains smooth because its vanishing Bloch radius compensates for the jump in $\hat{\boldsymbol n}$. Indeed, the Bloch vector
\begin{equation}
\boldsymbol r(t)
=
\tanh[\beta d(t)]\,\hat{\boldsymbol n}(t)
=
\frac{\tanh[\beta d(t)]}{d(t)}\,\boldsymbol d(t)
\simeq
\beta\boldsymbol v s,
\end{equation}
passes smoothly through the center of the Bloch ball, where $\rho(t_0)=\tfrac12\Id$, and traces a closed mixed-state loop. The two trajectories are illustrated in Fig.~\ref{fig:gap_paths}.

\subsection{Berry phase: a one-parameter family of regularized values}
\label{sec:gap-berry}

To assign a Berry phase, one must supplement the open path with a closure prescription. A natural regularization is to open the gap by adding a small constant bias field,
\begin{equation}
\boldsymbol{d}_\epsilon(t)=\boldsymbol{d}(t)+\epsilon\,\hat{\boldsymbol{u}},
\qquad d_\epsilon(t)\equiv|\boldsymbol{d}_\epsilon(t)|,\quad
\epsilon>0,\quad \hat{\boldsymbol{u}}\nparallel\hat{\boldsymbol{v}},
\label{eq:regularization}
\end{equation}
which keeps the gap open at all times, so that $\hat{\boldsymbol{n}}_\epsilon=\boldsymbol{d}_\epsilon/d_\epsilon$ is a smooth closed loop with Berry phase $\gamma_B(\epsilon\hat{\boldsymbol{u}})=-\frac12\Omega(\epsilon\hat{\boldsymbol{u}})$. As $\epsilon\to0$ the regularized Berry phase converges to a value that depends on the displacement direction through its azimuth $\alpha$~\cite{YangNoh2026geodesic}:
\begin{equation}
\gamma_B(\epsilon\hat{\boldsymbol{u}})
=-\tfrac12\,\Omega[C]-\alpha+\mathcal O(\epsilon),
\qquad
\alpha = \operatorname{atan2}\!\left(
\boldsymbol u_\perp\cdot\hat{\boldsymbol b},
\boldsymbol u_\perp\cdot\hat{\boldsymbol a}_\perp
\right).
\label{eq:berryfamily}
\end{equation}
Here $\boldsymbol u_\perp=\hat{\boldsymbol u}-(\hat{\boldsymbol u}\cdot\hat{\boldsymbol v})\hat{\boldsymbol v}$ is the component of the bias direction perpendicular to $\hat{\boldsymbol v}$. Let $\boldsymbol a=\ddot{\boldsymbol d}(t_0)$ and $\boldsymbol a_\perp=\boldsymbol a-(\boldsymbol a\cdot\hat{\boldsymbol v})\hat{\boldsymbol v}$, with $\hat{\boldsymbol a}_\perp=\boldsymbol a_\perp/|\boldsymbol a_\perp|$. For the noninflectional crossings considered here, $\boldsymbol a_\perp\neq0$, and $\{\hat{\boldsymbol v},\hat{\boldsymbol a}_\perp,\hat{\boldsymbol b}\}$, where $\hat{\boldsymbol b}=\hat{\boldsymbol v}\times\hat{\boldsymbol a}_\perp$, is the Frenet frame at the crossing. The angle $\alpha$ is the azimuth of $\boldsymbol u_\perp$ in the plane perpendicular to $\hat{\boldsymbol v}$, measured from the principal normal $\hat{\boldsymbol a}_\perp$ toward the binormal $\hat{\boldsymbol b}$. In the limit $\epsilon\to0$, the azimuth $\alpha$ determines the great-circle arc through $\pm\hat{\boldsymbol v}$ that closes the open path: $\alpha=0$ gives the geodesic in the osculating plane $\operatorname{span}\{\boldsymbol v,\boldsymbol a\}$, whereas $\alpha\neq0$ gives the geodesic obtained by rotating this plane through $\alpha$ about $\hat{\boldsymbol v}$; see Fig.~\ref{fig:gap_paths}(a).
The $\epsilon\to0$ limit therefore defines not a single Berry phase but a one-parameter family, $-\frac12\Omega[C]-\alpha$ with $\alpha\in(-\pi,\pi]$, one value per closing geodesic, and pure-state kinematics supplies no criterion that singles one out~\cite{GarzaSoto2023}.

The $\alpha$-independent part of Eq.~\eqref{eq:berryfamily} is intrinsic to the open curve. With the base point at the starting endpoint, $\hat{\boldsymbol n}_0=+\hat{\boldsymbol v}$, it is defined by the improper line integral
\begin{equation}
\Omega[C]\equiv\int_C\frac{\hat{\boldsymbol{n}}_0\cdot\big(\hat{\boldsymbol{n}}\times\exd\hat{\boldsymbol{n}}\big)}{1+\hat{\boldsymbol{n}}_0\cdot\hat{\boldsymbol{n}}}.
\label{eq:OmegaC}
\end{equation}
Note that the denominator vanishes at the antipode $\hat{\boldsymbol n}=-\hat{\boldsymbol n}_0=-\hat{\boldsymbol v}$. This is a gauge singularity of the monopole one-form, conventionally called a Dirac string; on $S^2$, the string appears as a puncture at the antipode and has no physical significance. The improper integral nevertheless converges because a window $|s|<w$ around the crossing contributes only $\mathcal O(w)$. For a closed curve that avoids the Dirac string, Eq.~\eqref{eq:OmegaC} gives the usual enclosed solid angle modulo $4\pi$. Along a great-circle arc joining $\pm\hat{\boldsymbol v}$, the azimuth is constant everywhere except at the singular antipodal endpoint, so the line integral over the nonsingular portion of the arc vanishes.

The $\alpha$-dependent offset arises from how the regularized loop passes the gauge singularity. Different closing arcs approach the antipode with different azimuths and therefore carry different gauge-transition contributions. The azimuthal turn imposed by the bias occurs once near each endpoint, but the monopole weight $1-\hat{\boldsymbol n}_0\cdot\hat{\boldsymbol n}$ approaches $2$ at the antipode and is only $\mathcal O(\epsilon)$ at the base point. The turn is therefore counted asymmetrically, producing the additional solid angle $2\alpha$ and hence the offset $-\alpha$ in Eq.~\eqref{eq:berryfamily}. The convergence of the improper integral and the origin of this closure-dependent term are derived in Appendix~\ref{app:gapproofs}.

Although the individual limiting Berry phase generally depends on the closing geodesic, two unambiguous predictions remain: a universal phase difference for opposite bias directions and, for a reflection-symmetric planar drive, a symmetry-selected pair of normal-bias phases. First, for any fixed transverse direction $\hat{\boldsymbol u}$, opposite biases give the universal phase difference $\gamma_B(+\epsilon\hat{\boldsymbol u})-\gamma_B(-\epsilon\hat{\boldsymbol u})\to\pi\;(\mathrm{mod}\;2\pi)$, because reversing the bias changes the closure azimuth by $\pi$. This is the familiar $\pi$ jump across a diabolical point, analyzed in Ref.~\cite{RakhechaWagh1996} and observed in neutron and atom interferometry~\cite{Wagh1998,Zhou2020}. Second, if the drive is confined to a plane through the origin, reflection symmetry singles out the two regularizations along opposite directions normal to that plane. For these two normal-bias limits, $\Omega[C]=0$ and $\alpha=\pm\pi/2$, giving $\gamma_B(\pm\epsilon\hat{\boldsymbol u})=\mp\pi/2$, with the sign fixed by $\operatorname{sgn}(\hat{\boldsymbol u}\cdot\hat{\boldsymbol b})$. Thus, the limit $\epsilon\to0$ exists for every fixed $\hat{\boldsymbol u}\nparallel\hat{\boldsymbol v}$, but its value generally depends on $\hat{\boldsymbol u}$; the reflection-symmetric case is special because symmetry identifies a canonical pair of normal-bias regularizations.

\subsection{Uhlmann phase across the crossing}
\label{sec:gap-uhlmann}

At every finite bath temperature $T_B>0$, the gap-closing Gibbs loop remains full rank even at $t_0$, so the Uhlmann construction remains applicable without introducing a gap-opening regularization. Two technical points nevertheless require care. First, unlike the conical loop of Sec.~III, the Bloch radius
$r(t)=\tanh[\beta d(t)]$ is no longer constant. Nevertheless, the terms in $\exd\sqrt{\rho}$ generated by $\exd r$ are diagonal in the instantaneous eigenbasis and therefore commute with $\sqrt{\rho}$. They do not contribute to $[\exd\sqrt{\rho},\sqrt{\rho}]$, so Eq.~\eqref{eq:AU_final_copy} continues to hold for $t\neq t_0$, with $r=r(t)$. As shown below, this connection admits a continuous extension through the crossing at every finite temperature.
Second, defining the phase requires choosing a reference time at which one traversal of the closed density-matrix path begins and ends. The crossing time $t_0$ is unsuitable for this purpose because $\rho(t_0)=\frac12\Id$, giving $\operatorname{Tr}[\rho(t_0)U_C]=\frac12\operatorname{Tr}U_C$. Since $U_C\in SU(2)$, this quantity is real, so its argument is restricted to $0$ or $\pi$ and is undefined when $\operatorname{Tr}U_C=0$. We therefore choose a fixed reference time $t_r\neq t_0$ and define $\gamma_U=\Argop\operatorname{Tr}[\rho(t_r)U_C(t_r)]$.

The resulting holonomy $U_C$ depends continuously on $T_B$. For the fixed reference time $t_r$, the associated phase is continuous wherever $\operatorname{Tr}[\rho(t_r)U_C(t_r)]\neq0$, and along the low-temperature branch considered here,
\begin{equation}
\lim_{T_B\to0^+}\gamma_U=-\frac12\Omega[C]\pmod{2\pi}.
\label{eq:uhlmann-unique}
\end{equation}
Proofs of these statements are given in Appendix~\ref{app:gapproofs}.

The mechanism is a mixed-state regularization of the diabolical point. Writing $\hat{\boldsymbol{n}}(t)={\rm sgn}(s)\,\hat{\boldsymbol{w}}(t)$, where
$\hat{\boldsymbol{w}}\propto\boldsymbol{d}(t)/s$ is smooth across the crossing, the antipodal jump of $\hat{\boldsymbol{n}}$ drops out of the connection: the sign enters $\hat{\boldsymbol{n}}\times\exd\hat{\boldsymbol{n}}$ twice and cancels, leaving $\hat{\boldsymbol{n}}\times\exd\hat{\boldsymbol{n}}=\hat{\boldsymbol{w}}\times\exd\hat{\boldsymbol{w}}$.
The only remaining factor in $A_U$ that could be singular is the weight $f(r)=1-\sqrt{1-r^2}$, and it is not: near $t_0$ the Gibbs radius is $r\simeq\beta v|s|$, so $f\simeq\frac12\beta^2v^2s^2$ vanishes quadratically,
and the transport switches itself off precisely where the pure-state
description fails. Geometrically, the Bloch vector traverses the straight segment through the center, which subtends no solid angle. As $T_B\to0$ the weight approaches unity at every fixed distance from the crossing, while the crossing window continues to contribute nothing, and Eq.~\eqref{eq:uhlmann-unique} follows.

The equatorial transition of Sec.~III provides a useful contrast. There, the phase jumps by $\pi$ at $T_C\simeq0.759$ (see Fig.~\ref{fig:Uhlmann_Phase}). For a simple gap-closing drive confined to a plane through the origin, with $\hat{\boldsymbol n}(t)$ traversing a single great-circle semicircle without backtracking, Appendix~\ref{app:gapproofs} shows that the Uhlmann overlap remains positive at every finite temperature. Consequently, the equatorial transition is removed and $\gamma_U=0$ for all $T_B>0$. This conclusion need not hold for planar paths with backtracking or additional winding, which can exhibit a $0$-to-$\pi$ phase jump. For such a simple planar drive, which is reflection-symmetric, the two normal-bias regularizations give the Berry values $\mp\pi/2$, whereas the Uhlmann phase has the single limiting value $\lim_{T_B\to0^+}\gamma_U=0=-\frac12\Omega[C]$.

\subsection{Geodesic selection rule}
\label{sec:gap-selection}

Combining Eqs.~\eqref{eq:berryfamily} and \eqref{eq:uhlmann-unique} identifies the fate of the Uhlmann--Berry correspondence at the degeneracy:
\begin{equation}
\lim_{T_B\to0^+}\gamma_U = 
\left.\lim_{\epsilon\to0^+}\gamma_B(\epsilon\hat{\boldsymbol u})\right|_{\alpha=0},
\qquad
\lim_{T_B\to0^+}\gamma_U - \lim_{\epsilon\to0^+}\gamma_B(\epsilon\hat{\boldsymbol u}) \equiv \alpha \pmod{2\pi}.
\label{eq:selection}
\end{equation}
The low-temperature limit of the mixed-state construction selects exactly one member of the Berry family, namely the osculating-plane closure $\alpha=0$. The Uhlmann--Berry correspondence established in the gapped case therefore extends to the gap closing in the form of a \emph{geodesic selection rule}. The geometric origin of this selection follows from the local expansion $\boldsymbol d(s)\simeq\boldsymbol v s+\frac12\boldsymbol a s^2$, where $\boldsymbol a=\ddot{\boldsymbol d}(t_0)$. Its transverse component is $\frac12\boldsymbol a_\perp s^2$, which points along $+\hat{\boldsymbol a}_\perp$ for both signs of $s$. Thus, both branches of the control curve lie on the same side of the tangent line near the degeneracy. The smooth Gibbs trajectory through the center of the Bloch ball inherits this local approach geometry and consequently selects the closing geodesic in the osculating plane $\operatorname{span}\{\boldsymbol v,\boldsymbol a\}$. This realizes, in closed form, the closing prescription proposed operationally in Ref.~\cite{GarzaSoto2023}. For the noninflectional transversal crossings considered here, we therefore use $\gamma_B[C]\equiv-\frac12\Omega[C]\;(\mathrm{mod}\;2\pi)$ to denote the Berry value selected by the osculating-plane regularization, $\alpha=0$. 
This result should be distinguished from the failure of the zero-temperature correspondence found for systems with a degenerate ground-state manifold, where the Uhlmann phase need not coincide with the scalar Wilczek--Zee phase~\cite{Wang2025degenerate}. In the present problem, the spectrum is nondegenerate away from a single isolated crossing; the ambiguity instead lies in completing the resulting open pure-state path, and the pure-state limit of the Uhlmann construction selects the osculating-plane completion. 
Table~\ref{tab:gapclosing} summarizes the comparison.

\begin{table}
\caption{Berry and Uhlmann phases at a transversal gap-closing degeneracy. $C$ is the open curve traced by $\hat{\boldsymbol n}$, $\Omega[C]$ is the open-path contribution defined in Eq.~\eqref{eq:OmegaC}, and $\alpha$ is the closure azimuth.}
\label{tab:gapclosing}
\begin{ruledtabular}
\begin{tabular}{lcc}
 & Berry (pure) & Uhlmann (Gibbs) \\
\colrule
State path & open $C\subset S^2$ & closed, through $\tfrac12\Id$ \\
Regularization & required & none \\
Phase & $-\tfrac12\Omega[C]-\alpha$ & unique when defined \\
Low-temperature limit & depends on $\hat{\boldsymbol u}$ & $-\tfrac12\Omega[C]$ \\
Simple planar crossing & $\mp\pi/2$ (normal bias) & $0$ for all $T_B>0$ \\
\end{tabular}
\end{ruledtabular}
\end{table}

\begin{figure}
    \centering
    \includegraphics[width=0.8\linewidth]{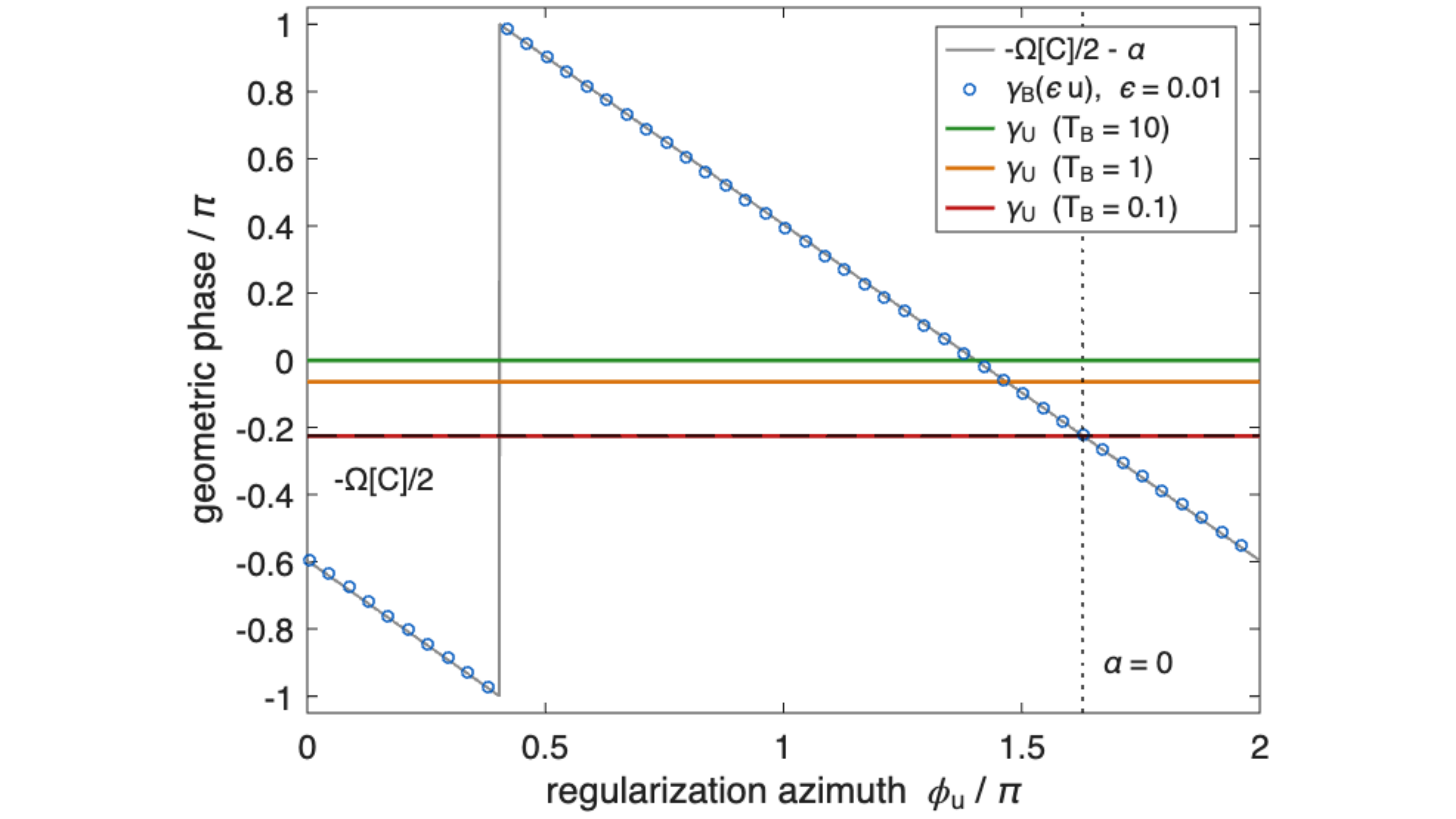}
    \caption{Geodesic selection rule for the drive of Eq.~\eqref{eq:demo-field}. Circles: regularized Berry phase $\gamma_B(\epsilon\hat{\boldsymbol{u}})$ at $\epsilon=0.01$ versus the azimuth $\phi_u$ of the bias direction; the gray line is $-\frac12\Omega[C]-\alpha(\phi_u)$ with $\Omega[C]=0.4485\pi$. Horizontal lines: Uhlmann phase at $T_B=10,\,1,\,0.1$ ($\gamma_U/\pi=-0.0003~\text{(green)},\,-0.0639~\text{(orange)},\,-0.2254~\text{(red)}$); dashed line: the pure-state limit $-\frac12\Omega[C]=-0.2243\pi$. The intersection occurs at the azimuth $\phi_{\hat{\boldsymbol{a}}_\perp}=1.628\pi$ of the principal normal (dotted vertical line), agreeing with the osculating-plane closure, $\alpha=0$.}
    \label{fig:gap_phases}
\end{figure}

Figure~\ref{fig:gap_phases} demonstrates the rule numerically for the driving field
\begin{equation}
\boldsymbol{d}(t)=\mathfrak{Q}\begin{pmatrix}\sin t\\[1pt] 0.8\sin2t+0.3\cos3t-0.3\\[1pt] 1-\cos t\end{pmatrix}, \quad 
\label{eq:demo-field}
\end{equation}
where
\begin{align}
    \mathfrak Q=\left(\begin{matrix}0.719101&-0.449438&-0.529999\\-0.449438&0.280899&-0.847998\\0.529999&0.847998&0\end{matrix}\right).
\end{align}
The derivative of the unrotated curve at $t_0=0$ is proportional to $(1,1.6,0)$, and $\mathfrak Q$ maps this vector onto $\hat{\boldsymbol z}$. Hence the crossing velocity satisfies $\dot{\boldsymbol d}(0)\parallel\hat{\boldsymbol z}$, and the open path has endpoints $\pm\hat{\boldsymbol z}$. The rotation $\mathfrak Q$ is chosen so that the laboratory $x$ and $y$ axes have no special relation to the Frenet frame. In particular, the principal-normal azimuth $\phi_{\hat{\boldsymbol a}_\perp}=1.628\pi$ is not used as an input in computing the Berry and Uhlmann curves but is determined independently from the first two derivatives of $\boldsymbol d(t)$ at the crossing. As the bias azimuth $\phi_u$ is varied, the regularized Berry phase follows the line $-\frac12\Omega[C]-\alpha(\phi_u)$, with slope $-1$ on an unwrapped phase branch and a $2\pi$ winding over one revolution. Because the Uhlmann construction requires no bias, its phases at $T_B=10,\,1,\,0.1$ are independent of $\phi_u$ and appear as horizontal lines that approach $\gamma_B[C]=-\frac12\Omega[C]=-0.2243\pi$. The Berry line intersects this limiting value at the azimuth $\phi_u=\phi_{\hat{\boldsymbol a}_\perp}$, independently reproducing the principal-normal direction and thereby verifying Eq.~\eqref{eq:selection}.

\section{Conclusion}
\label{sec:conclusion}

We have followed Berry and Uhlmann holonomies for the same driven-qubit Hamiltonian from gapped conical cycles to an isolated gap closing. For the gapped cycles, both holonomies were obtained in closed form and compared as functions of bath temperature and the control-field parameters, including the discontinuous equatorial $\pi$ jump. The open-system analysis then supplied the dynamical foundation of this comparison. The autonomous rotating-frame Lindblad equation has a unique fixed point, which corresponds in the laboratory frame to a periodic limit cycle rather than to the instantaneous Gibbs loop. Its lag corrections scale as positive powers of the adiabaticity parameter $\eta=\omega/\omega_0$ and vanish as $\eta\to0$, so the Gibbs loop is recovered in the adiabatic limit. Taking this limit first makes the intermediate equilibrium loop explicit; subsequent cooling turns the Gibbs state into the rank-one ground-state projector, and the Uhlmann transport projected onto this one-dimensional support reduces to the Berry $U(1)$ holonomy. Under the assumptions used here, however, the adiabatic and zero-temperature limits commute, and their joint limit gives the Berry phase. The route through the equilibrium loop may be written as $\gamma_U^{\rm dyn}\xrightarrow{\eta\to0}\gamma_U^{\rm eq}\xrightarrow{T_B\to0}\gamma_B$. Both limits require a gap that remains open throughout the cycle.

We have followed Berry and Uhlmann holonomies for the same driven-qubit Hamiltonian from gapped conical cycles to an isolated gap closing. For the gapped cycles, both holonomies were obtained in closed form and compared as functions of bath temperature and the control-field parameters, including the discontinuous equatorial $\pi$ jump. The open-system analysis then supplied the dynamical foundation of this comparison. In the rotating frame, the driven Lindblad equation has a unique fixed point, which corresponds in the laboratory frame to a periodic limit cycle rather than to the instantaneous Gibbs loop. Its lag corrections scale as positive powers of the adiabaticity parameter $\eta=\omega/\omega_0$ and vanish as $\eta\to0$, so the Gibbs loop is recovered as the adiabatic fixed point of the dynamics. Cooling this equilibrium loop subsequently turns the Gibbs state into the rank-one ground-state projector, and the Uhlmann transport projected onto this one-dimensional support reduces to the Berry $U(1)$ holonomy. Together, these limits form the ordered hierarchy $\gamma_U^{\rm dyn}\xrightarrow{\eta\to0}\gamma_U^{\rm eq} \xrightarrow{T_B\to0}\gamma_B$. Both steps, however, require a gap that remains open throughout the cycle.

At an isolated noninflectional transversal crossing, the two constructions separate. The pure-state trajectory becomes an open curve with antipodal endpoints, and regularization generates the one-parameter Berry family $-\frac12\Omega[C]-\alpha$, while pure-state kinematics supplies no criterion for selecting one of its members. The Gibbs trajectory instead passes smoothly through the maximally mixed state. For every finite $T_B>0$, the Uhlmann connection and holonomy are defined without introducing a gap-opening regularization and vary continuously with temperature; along the low-temperature branch considered here, the associated phase approaches $-\frac12\Omega[C]$. The reason is local: at the crossing the Gibbs state becomes maximally mixed, so the Uhlmann connection loses its sensitivity to the Bloch direction that is undefined in the pure-state description. Mathematically, its state-dependent weight vanishes as $f(r)\simeq r^2/2$ when $r\to0$, allowing the connection to extend smoothly through the degeneracy. These results yield the central conclusion of this work. The low-temperature Uhlmann construction does not reproduce the Berry family as a whole but selects the single value $\gamma_B[C]=-\frac12\Omega[C]$, corresponding to the osculating-plane closure $\alpha=0$. The selection is geometric. Near the crossing, both branches of the control curve bend to the same side of the tangent line, thereby defining a unique osculating plane. As the Gibbs trajectory passes smoothly through the maximally mixed state, it carries this local curvature information across the crossing and selects the geodesic closure in that plane. The Uhlmann--Berry correspondence therefore survives the gap closing as a geodesic selection rule.

Our study presents two potential directions for future research. First, the principal dynamical limitation of the present treatment is the crossing region itself. There the secular approximation fails and the passage becomes a dissipative Landau--Zener problem, whose geometric phase and interplay with the finite-frequency lags quantified in Sec.~\ref{sec:ss} require a nonsecular dynamical treatment. A related non-Hermitian problem is already present in the gapped rotating-frame generator $M$ of Eq.~\eqref{eq:Mmatrix}: its repeated-root locus is determined by the vanishing of the characteristic-polynomial discriminant derived in Appendix~\ref{app:eig}. Establishing where this locus is genuinely defective, and whether the resulting exceptional points leave distinct signatures in the lag or in the Uhlmann holonomy, is a complementary open problem.

Second, the same local geometry suggests an extension from a driving period to the Brillouin zone. For a one-dimensional two-band Bogoliubov--de Gennes (BdG) Hamiltonian $H(k)=-\boldsymbol d(k)\cdot\boldsymbol\sigma$, an isolated band closing has the same local form with the crystal momentum $k$ as the path parameter. In the Kitaev chain, the momentum-space Gibbs curve passes through the maximally mixed state at the topological transition \cite{Viyuela2014_1D,Andersson2016}. This suggests extending the present selection rule to the occupied-band Zak phase. Such an extension must incorporate particle--hole or chiral symmetry, Brillouin-zone periodicity, and possibly multiple symmetry-related crossings; the symmetry constraints on admissible gap-opening terms would then determine which closure azimuths are physically accessible.

Experimentally, the $\pi$ jump and the one-sided values $\mp\pi/2$ can be tested in neutron and atom interferometry \cite{Wagh1998,Zhou2020}. Superconducting-circuit protocols for measuring Uhlmann phases \cite{Viyuela2018Obs} could probe the regularity of the holonomy through a gap closing and, for the simple reflection-symmetric protocol analyzed here, the predicted disappearance of the equatorial $\pi$ jump.


\appendix

\section{The time-independent effective Hamiltonian in the rotating frame} \label{App1}

To remove the explicit time dependence in Eq.~\eqref{eq:H_lab}, we perform a rotation around the $z$ axis,
\begin{equation}
R(t)=\exp\!\left[-i\frac{\omega t}{2}\sigma_z\right],
\qquad
|\psi_{\mathrm{rot}}(t)\rangle = R^\dagger(t)\,|\psi(t)\rangle .
\label{eq:rot_frame_def}
\end{equation}
The rotating-frame Hamiltonian is
\begin{equation}
H_{\mathrm{eff}}
= R^\dagger H R - i\,R^\dagger \dot R
= -d_0\,\sigma_x - \Bigl(d_1+\frac{\omega}{2}\Bigr)\sigma_z,
\label{eq:Heff}
\end{equation}
which is time-independent.  Defining
\begin{equation}
\Delta \equiv d_1+\frac{\omega}{2},
\qquad
\boldsymbol{v}\equiv (d_0,0,\Delta),
\qquad
\OmR\equiv|\boldsymbol{v}|=\sqrt{d_0^2+\Delta^2},
\label{eq:vOmega}
\end{equation}
we can write $H_{\mathrm{eff}}=-\boldsymbol{v}\cdot\boldsymbol{\sigma}$.

Because $H_{\mathrm{eff}}$ is constant, the exact propagator can be expressed in closed form:
\begin{equation}
U(t)=R(t)\,e^{-iH_{\mathrm{eff}}t},
\qquad
e^{-iH_{\mathrm{eff}}t}
=\cos(\OmR t)\,\Id + i\sin(\OmR t)\,\frac{\boldsymbol{v}\cdot\boldsymbol{\sigma}}{\OmR}.
\label{eq:U_exact}
\end{equation}
At one period $t=T$, since $R(T)=\exp[-i\pi\sigma_z]=-\Id$, we have
\begin{equation}
U(T)= -e^{-iH_{\mathrm{eff}}T}.
\label{eq:UT}
\end{equation}

\section{\texorpdfstring{Bloch precession and analytic $\phi_{\mathrm{dyn}}$}{Bloch precession and analytic dynamical phase}}

In the rotating frame the Hamiltonian is constant, $H_{\mathrm{eff}}=-\boldsymbol{v}\cdot\boldsymbol{\sigma}$, and the Bloch vector obeys
\begin{equation}
\frac{\exd\boldsymbol{r}_{\mathrm{rot}}}{\exd t}
=2\,\boldsymbol{r}_{\mathrm{rot}}\times\boldsymbol{v},
\label{eq:bloch_eq}
\end{equation}
which describes a rigid precession of $\boldsymbol{r}_{\mathrm{rot}}$ about the fixed axis $\boldsymbol{v}$ with angular speed $2\OmR$.
Introducing the unit vector $\hat{\boldsymbol{v}}=\boldsymbol{v}/\OmR$ and the scalar
\begin{equation}
c \equiv \hat{\boldsymbol{v}}\cdot\boldsymbol{n}_0
=\frac{\boldsymbol{v}\cdot\boldsymbol{n}_0}{\OmR},
\label{eq:c_def}
\end{equation}
Decomposing $\boldsymbol{n}_0$ into components parallel/perpendicular to $\hat{\boldsymbol{v}}$,
$\boldsymbol{n}_{0,\parallel}=(\boldsymbol{n}_0\!\cdot\!\hat{\boldsymbol{v}})\hat{\boldsymbol{v}}$ and $\boldsymbol{n}_{0,\perp}=\boldsymbol{n}_0-\boldsymbol{n}_{0,\parallel}$, the parallel component is conserved while the perpendicular component rotates in the transverse plane.  With $c\equiv \boldsymbol{n}_0\!\cdot\!\hat{\boldsymbol{v}}$ and $\boldsymbol{r}_{\mathrm{rot}}(0)=\boldsymbol{n}_0$, this immediately gives
\begin{equation}
\boldsymbol{n}_0\cdot\boldsymbol{r}_{\mathrm{rot}}(t)
=|\boldsymbol{n}_{0,\parallel}|^2+|\boldsymbol{n}_{0,\perp}|^2\cos(2\OmR t)
=c^2+(1-c^2)\cos(2\OmR t),
\label{eq:n0_rrot}
\end{equation}
where $|\boldsymbol{n}_0|=1$ for an initial pure state.

Combining Eqs.~\eqref{eq:phi_dyn_def}, and \eqref{eq:n0_rrot}, the dynamical phase becomes
\begin{equation}
\phi_{\mathrm{dyn}}
= d\int_0^T \exd t\,\Bigl[c^2+(1-c^2)\cos(2\OmR t)\Bigr]
= d\left[c^2 T + (1-c^2)\frac{\sin(2\OmR T)}{2\OmR}\right].
\label{eq:phi_dyn_closed}
\end{equation}
For the present model, $c$ can be written explicitly as
\begin{equation}
c=\frac{d_0^2+\Delta d_1}{\OmR\,d}.
\label{eq:c_explicit}
\end{equation}

\section{Detailed Derivation of the Uhlmann connection}
\label{App:UhlCon}

\subsection{\texorpdfstring{Compute $[\exd\sqrt{\rho},\sqrt{\rho}]$}{Evaluation of the square-root commutator}}
From the spectral decomposition,
\begin{equation}
\rho=\lambda_+P_+ + \lambda_-P_-,
\qquad
\sqrt{\rho}=\sqrt{\lambda_+}\,P_+ + \sqrt{\lambda_-}\,P_-.
\label{eq:sqrt_rho_copy}
\end{equation}
Since $\lambda_\pm$ are constants along the loop,
\begin{equation}
\exd\sqrt{\rho}=\sqrt{\lambda_+}\,\exd P_+ + \sqrt{\lambda_-}\,\exd P_-.
\label{eq:dsqrt1_copy}
\end{equation}
Using $P_+ + P_-=\Id$, we have $\exd P_-=-\exd P_+$ and thus
\begin{equation}
\exd\sqrt{\rho}=\big(\sqrt{\lambda_+}-\sqrt{\lambda_-}\big)\,\exd P_+.
\label{eq:dsqrt2_copy}
\end{equation}
Therefore,
\begin{align}
\big[\exd\sqrt{\rho},\sqrt{\rho}\big]
&=\big(\sqrt{\lambda_+}-\sqrt{\lambda_-}\big)\Big[\exd P_+,\ \sqrt{\lambda_+}P_+ + \sqrt{\lambda_-}P_-\Big]\nonumber\\
&=\big(\sqrt{\lambda_+}-\sqrt{\lambda_-}\big)^2\,[\exd P_+,P_+].
\label{eq:comm_reduction_copy}
\end{align}
Define
\begin{equation}
f(r)\equiv\big(\sqrt{\lambda_+}-\sqrt{\lambda_-}\big)^2
=1-\sqrt{1-r^2}.
\label{eq:fr_copy}
\end{equation}
Next, using $P_+=\frac{1}{2}(\Id+\hat{\boldsymbol{n}}\cdot\boldsymbol{\sigma})$,
\begin{equation}
\exd P_+=\frac{1}{2}\big(\exd\hat{\boldsymbol{n}}\cdot\boldsymbol{\sigma}\big).
\label{eq:dP_copy}
\end{equation}
Then
\begin{equation}
[\exd P_+,P_+]
=\frac{1}{4}\Big[\exd\hat{\boldsymbol{n}}\cdot\boldsymbol{\sigma},\ \hat{\boldsymbol{n}}\cdot\boldsymbol{\sigma}\Big]
=\frac{i}{2}\,(\exd\hat{\boldsymbol{n}}\times\hat{\boldsymbol{n}})\cdot\boldsymbol{\sigma}
=-\frac{i}{2}\,(\hat{\boldsymbol{n}}\times \exd\hat{\boldsymbol{n}})\cdot\boldsymbol{\sigma},
\label{eq:commP_copy}
\end{equation}
where we used the Pauli identity
\begin{equation}
[\boldsymbol{a}\cdot\boldsymbol{\sigma},\boldsymbol{b}\cdot\boldsymbol{\sigma}]
=2i(\boldsymbol{a}\times\boldsymbol{b})\cdot\boldsymbol{\sigma}.
\label{eq:pauli_comm_copy}
\end{equation}
Combining Eqs.~\eqref{eq:comm_reduction_copy} and \eqref{eq:commP_copy}, we obtain
\begin{equation}
\big[\exd\sqrt{\rho},\sqrt{\rho}\big]
=-\frac{i}{2}\,f(r)\,(\hat{\boldsymbol{n}}\times \exd\hat{\boldsymbol{n}})\cdot\boldsymbol{\sigma}.
\label{eq:rhs_copy}
\end{equation}

\subsection{\texorpdfstring{Solve $\{\rho,A_U\}=[\exd\sqrt{\rho},\sqrt{\rho}]$}{Solution of the parallel-transport equation}}
Any traceless anti-Hermitian $2\times2$ matrix can be parameterized as
\begin{equation}
A_U=-\frac{i}{2}\,\boldsymbol{a}\cdot\boldsymbol{\sigma},\qquad \boldsymbol{a}\in\mathbb{R}^3.
\label{eq:AU_ansatz_copy}
\end{equation}
Using $\rho=\frac{1}{2}(\Id+r\,\hat{\boldsymbol{n}}\cdot\boldsymbol{\sigma})$, the left-hand side becomes
\begin{align}
\rho A_U + A_U\rho
&=-\frac{i}{4}\Big\{\Id+r\,\hat{\boldsymbol{n}}\cdot\boldsymbol{\sigma},\ \boldsymbol{a}\cdot\boldsymbol{\sigma}\Big\}\nonumber\\
&=-\frac{i}{4}\Big(2\,\boldsymbol{a}\cdot\boldsymbol{\sigma}+2r(\hat{\boldsymbol{n}}\cdot\boldsymbol{a})\,\Id\Big)\nonumber\\
&=-\frac{i}{2}\,\boldsymbol{a}\cdot\boldsymbol{\sigma}-\frac{i}{2}\,r(\hat{\boldsymbol{n}}\cdot\boldsymbol{a})\,\Id,
\label{eq:LHS_copy}
\end{align}
where we used $\{\Id,\boldsymbol{a}\cdot\boldsymbol{\sigma}\}=2\boldsymbol{a}\cdot\boldsymbol{\sigma}$ and
$\{\hat{\boldsymbol{n}}\cdot\boldsymbol{\sigma},\boldsymbol{a}\cdot\boldsymbol{\sigma}\}=2(\hat{\boldsymbol{n}}\cdot\boldsymbol{a})\Id$.
Since the right-hand side \eqref{eq:rhs_copy} is traceless, the identity component on the left must vanish, which enforces
\begin{equation}
\hat{\boldsymbol{n}}\cdot\boldsymbol{a}=0.
\label{eq:orth_copy}
\end{equation}
Under \eqref{eq:orth_copy}, Eq.~\eqref{eq:LHS_copy} reduces to
\begin{equation}
\rho A_U + A_U\rho=-\frac{i}{2}\,\boldsymbol{a}\cdot\boldsymbol{\sigma}.
\label{eq:LHS_simpl_copy}
\end{equation}
Equating \eqref{eq:LHS_simpl_copy} with \eqref{eq:rhs_copy} yields
\begin{equation}
\boldsymbol{a}=f(r)\,(\hat{\boldsymbol{n}}\times \exd\hat{\boldsymbol{n}}),
\label{eq:a_sol_copy}
\end{equation}

\section{Detailed Derivation of the Uhlmann holonomy}
\label{App:UhlHol}

By definition, $U(\phi)$ satisfies the following first-order linear differential equation with the initial condition $U(0)=\Id$:
\begin{equation}
\frac{\exd U(\phi)}{\exd\phi} = A_U(\phi) U(\phi) = -\frac{i}{2}f(r)\Big[-\cos\theta\sin\theta(\cos\phi\,\sigma_x+\sin\phi\,\sigma_y)+\sin^2\theta\sigma_z\Big] U(\phi)
\end{equation}
To eliminate the explicit $\phi$-dependence, we move to a rotating frame by introducing a unitary transformation $R(\phi)$:
\begin{equation}
R(\phi) = \exp\left(-i \frac{\phi}{2} \sigma_z\right)
\end{equation}
We define a new evolution operator $\tilde{U}(\phi)$ such that $U(\phi) = R(\phi) \tilde{U}(\phi)$. 
Using the relation $R(\phi) \sigma_x R^\dagger(\phi) = \cos\phi\,\sigma_x + \sin\phi\,\sigma_y$, we can rewrite $A_U(\phi)$ as:
\begin{equation}
A_U(\phi) = -\frac{i}{2}f(r)\Big[ -\cos\theta\sin\theta R(\phi) \sigma_x R^\dagger(\phi) + \sin^2\theta \sigma_z \Big]
\end{equation}
Substituting $U(\phi) = R(\phi) \tilde{U}(\phi)$ into the differential equation yields:
\begin{equation}
\frac{\exd U(\phi)}{\exd\phi}=\frac{\exd R(\phi)}{\exd\phi} \tilde{U}(\phi) + R(\phi) \frac{\exd\tilde{U}(\phi)}{\exd\phi} = A_U(\phi) R(\phi) \tilde{U}(\phi)
\end{equation}
Since $\frac{\exd R}{\exd\phi} = -\frac{i}{2}\sigma_z R$, we have:
\begin{equation}
-\frac{i}{2}\sigma_z R \tilde{U} + R \frac{\exd\tilde{U}}{\exd\phi} = A_U R \tilde{U}
\end{equation}
Multiplying by $R^\dagger(\phi)$ from the left, we obtain a differential equation for $\tilde{U}(\phi)$ with constant coefficients:
\begin{equation}
\frac{\exd\tilde{U}(\phi)}{\exd\phi} = R^\dagger \left( A_U + \frac{i}{2}\sigma_z \right) R \tilde{U} = \frac{i}{2} \Big[ f(r)\cos\theta\sin\theta \sigma_x + (1-f(r)\sin^2\theta )\sigma_z \Big] \tilde{U}(\phi)
\end{equation}
We define the norm of the effective magnetic field vector $\Theta$:
\begin{align}
\Theta & = \sqrt{ (f(r)\cos\theta\sin\theta)^2 + (1-f(r)\sin^2\theta )^2 }\nonumber \\
& = \sqrt{\sin^2\theta \big( f(r)^2 - 2f(r) \big) + 1}
\label{eqn:Theta}
\end{align}
Since the matrix is now independent of $\phi$, we can directly integrate it to find $\tilde{U}(\phi)$. Using Euler's formula for Pauli matrices ($\exp(i\alpha \hat{n}\cdot\boldsymbol{\sigma}) = \cos\alpha \Id + i\sin\alpha (\hat{n}\cdot\boldsymbol{\sigma})$), we get:
\begin{equation}
\tilde{U}(\phi) = \cos\left(\frac{\phi\Theta}{2}\right) \Id + i \frac{\sin\left(\frac{\phi\Theta}{2}\right)}{\Theta} \Big[ f(r)\cos\theta\sin\theta \sigma_x + (1-f(r)\sin^2\theta )\sigma_z \Big]
\end{equation}
Finally, the original operator $U(\phi)$ is given by:
\begin{equation}
U(\phi) = \exp\left(-i \frac{\phi}{2} \sigma_z\right) \tilde{U}(\phi)
\end{equation}
For a full period $T$ (where $\phi = 2\pi$), the rotation operator becomes $R(2\pi) = \exp(-i \pi \sigma_z) = -\Id$.

\section{Bloch equation and relaxation tensor}
\label{app:bloch}

We derive Eq.~\eqref{eq:relax} by mapping the dissipator of Eq.~\eqref{eq:gksl_general} onto the Bloch
vector in the field-adapted basis, where $H=-d\,\sigma_3$, $|g\rangle$ is the $+1$ eigenstate of
$\sigma_3$, and $\rho=\tfrac12(\Id+\boldsymbol{r}\cdot{\boldsymbol{\sigma}})$.

For the emission, with $L_\downarrow=|g\rangle\langle e|$ and
$L_\downarrow^\dagger L_\downarrow=|e\rangle\langle e|$,
\begin{equation}
\begin{gathered}
\mathcal D[L_\downarrow]\rho=\begin{pmatrix}\tfrac{1-r_3}{2}&-\tfrac{r_1-ir_2}{4}\\[2pt]
-\tfrac{r_1+ir_2}{4}&-\tfrac{1-r_3}{2}\end{pmatrix},\\[3pt]
\dot r_3=\gamma_\downarrow(1-r_3),\quad \dot r_{1,2}=-\tfrac{\gamma_\downarrow}{2}r_{1,2}.
\end{gathered}
\end{equation}
For the absorption, with $L_\uparrow=|e\rangle\langle g|$,
$\dot r_3=-\gamma_\uparrow(1+r_3)$, $\dot r_{1,2}=-\tfrac{\gamma_\uparrow}{2}r_{1,2}$.
\emph{Total.} Summing the two channels,
\begin{equation}
\dot r_3=-(\gamma_\downarrow+\gamma_\uparrow)(r_3-\req),\qquad
\dot r_{1,2}=-\tfrac{\gamma_\downarrow+\gamma_\uparrow}{2}\,r_{1,2},
\end{equation}
with $\req=(\gamma_\downarrow-\gamma_\uparrow)/(\gamma_\downarrow+\gamma_\uparrow)=(2\nth+1)^{-1}=\tanh(\beta d)$.
Thus the longitudinal rate is $1/T_1=\gamma_\downarrow+\gamma_\uparrow$ and the induced
transverse rate is half of it. \emph{Pure dephasing.} Using
$\sigma_z\sigma_{1,2}\sigma_z=-\sigma_{1,2}$,
$\tfrac{\gamma_\phi}{2}\mathcal D[\sigma_z]\rho$ gives $\dot r_{1,2}=-\gamma_\phi r_{1,2}$,
$\dot r_3=0$. \emph{Relaxation tensor.} Collecting all channels,
$\dot{\boldsymbol{r}}|_{\rm diss}=-\Gamma(\boldsymbol{r}-\boldsymbol{r}_{\rm eq})$ with
$\Gamma=\mathrm{diag}(1/T_2,1/T_2,1/T_1)$ and $1/T_2=1/(2T_1)+\gamma_\phi$, i.e.\
Eq.~\eqref{eq:relax}. Rotating the field-adapted frame into the laboratory frame gives the
time-dependent $\Geff_{\rm lab}(t)=\mathcal R(t)\Geff\mathcal R(t)^{\mathsf T}$ of Eq.~\eqref{eq:blocheq}.

\section{Rotating-frame eigenvalues and propagator}
\label{app:eig}

For $M$ in Eq.~\eqref{eq:Mmatrix},
\begin{equation}
\begin{aligned}
\det(\mu \Id-M)={}&\mu^3+(2a+b)\mu^2 +(a^2+2ab+|{\boldsymbol{\Omega}}|^2)\mu +(a^2b+a\Omega_1^2+b\Omega_3^2),
\end{aligned}
\end{equation}
with $|{\boldsymbol{\Omega}}|^2=\Omega_1^2+\Omega_3^2$; the constant term equals $-\det M=abD$. The shift $\mu=x-(2a+b)/3$ removes the quadratic term, giving $x^3+px+q=0$ with
\begin{equation}
\begin{gathered}
p=|{\boldsymbol{\Omega}}|^2-\frac{(a-b)^2}{3},\\[2pt]
q=\frac{(a-b)\big[9(\Omega_1^2-2\Omega_3^2)-2(a-b)^2\big]}{27}.
\end{gathered}
\end{equation}
The discriminant of the depressed cubic is $\Delta=-4p^{3}-27q^{2}$: for $\Delta<0$ there is one real root and a complex-conjugate pair, for $\Delta>0$ three real roots, and for $\Delta=0$ a repeated root. Since $q^{2}\ge0$, any $p>0$ gives $\Delta<0$, so the condition $p>0$, i.e.\ $\sqrt3\,|\boldsymbol{\Omega}|>|a-b|$, is sufficient but not necessary for the underdamped case. 
The real root is then given by Cardano's formula,
\begin{equation}
x_1=\sqrt[3]{-\tfrac q2+\sqrt{\tfrac{q^2}{4}+\tfrac{p^3}{27}}}
+\sqrt[3]{-\tfrac q2-\sqrt{\tfrac{q^2}{4}+\tfrac{p^3}{27}}}.
\end{equation}
Factoring $(x-x_1)$ out of the reduced cubic polynomial $x^3+px+q$ and restoring $\mu=x-(2a+b)/3$ yields the eigenvalues of Eq.~\eqref{eq:eig} with
\begin{equation}
\Gamma_L=\tfrac{2a+b}{3}-x_1,\quad
\Gamma_T=\tfrac{2a+b}{3}+\tfrac{x_1}{2},\quad
\Omega_{\rm eff}=\tfrac12\sqrt{3x_1^2+4p},
\end{equation}
obeying the trace rule $\Gamma_L+2\Gamma_T=2a+b$. Treating $\Geff$ as a first-order
perturbation of $[{\boldsymbol{\Omega}}]_\times$ (whose eigenvalues are $0,\pm i|{\boldsymbol{\Omega}}|$) gives the
weak-coupling forms
\begin{align}
\Gamma_L&\simeq\frac1{T_1}+\Big(\frac1{T_2}-\frac1{T_1}\Big)\frac{\Omega_1^2}{|{\boldsymbol{\Omega}}|^2},\notag\\
\Gamma_T&\simeq\frac1{T_2}-\frac12\Big(\frac1{T_2}-\frac1{T_1}\Big)\frac{\Omega_1^2}{|{\boldsymbol{\Omega}}|^2},
\qquad \Omega_{\rm eff}\simeq|{\boldsymbol{\Omega}}|,
\end{align}
which reduce to Eq.~\eqref{eq:eigadiab} when $\eta\ll1$.

With distinct eigenvalues, $e^{Mt}=\sum_k e^{\mu_k t}P_k$ and
$P_k=\prod_{j\neq k}(M-\mu_j\Id)/(\mu_k-\mu_j)$. Realizing the complex pair, the
longitudinal projector is
\begin{equation}
P_L=\frac{(M+\Gamma_T\Id)^2+\Omega_{\rm eff}^2\Id}{(\Gamma_T-\Gamma_L)^2+\Omega_{\rm eff}^2}.
\end{equation}
Setting $N\equiv(M+\Gamma_T\Id)(\Id-P_L)$, one has $N^2=-\Omega_{\rm eff}^2(\Id-P_L)$ from
Cayley--Hamilton on the invariant plane, so the exponential closes into trigonometric form and
\begin{align}
e^{Mt}={}&e^{-\Gamma_Lt}P_L
+e^{-\Gamma_Tt}\Big[\cos(\Omega_{\rm eff}t)(\Id-P_L)\notag\\
&+\frac{\sin(\Omega_{\rm eff}t)}{\Omega_{\rm eff}}(M+\Gamma_T\Id)(\Id-P_L)\Big].
\label{eq:propagator}
\end{align}
The state relaxes along $P_L$ at rate $\Gamma_L$ and spirals in the complementary plane at
frequency $\Omega_{\rm eff}$ with decay $\Gamma_T$. In the isotropic case $\gamma_\phi=1/(2T_1)$
($T_1=T_2$), Eq.~\eqref{eq:propagator} reduces to a Rodrigues rotation times uniform decay
$e^{-t/T_2}$; the ladder-only model has $T_2=2T_1$ (anisotropic), requiring the general form.

\section{Steady-state lag and adiabatic purity}
\label{app:ss}

Stationarity ${\boldsymbol{\Omega}}\times\boldsymbol{r}_{\rm ss}=\Gamma\boldsymbol{\delta}$ with
$\boldsymbol{r}_{\rm ss}=\req\hat{\boldsymbol{e}}_3+\boldsymbol{\delta}$ and $\Geff=\mathrm{diag}(1/T_2,1/T_2,1/T_1)$ reads
$-\Omega_3\delta_2=\delta_1/T_2$, $\Omega_3\delta_1-\Omega_1\delta_3-\req\Omega_1=\delta_2/T_2$,
$\Omega_1\delta_2=\delta_3/T_1$. Eliminating $\delta_{1,3}$ gives $\delta_2=-\req\Omega_1/D$ and
Eq.~\eqref{eq:lag}, with $D=1/T_2+\Omega_1^2T_1+\Omega_3^2T_2$. In control parameters,
$\Omega_1=\omega\sin\theta$, $\Omega_3\simeq-\omega_0$, $\req=(2\nth+1)^{-1}$, $1/T_1=\gamma(2\nth+1)$,
$1/T_2=1/(2T_1)+\gamma_\phi$, and $D\simeq\Omega_3^2T_2$, yielding the leading orders of
Eq.~\eqref{eq:lagadiab}.

The limit-cycle radius follows in closed form. Substituting Eq.~\eqref{eq:lag} into
$\rss^2=\delta_1^2+\delta_2^2+(\req+\delta_3)^2$ and writing
$K\equiv1/T_2+\Omega_3^2T_2$, so that $D=K+\Omega_1^2T_1$, the numerator collects into
$K^2+\Omega_1^2+\Omega_1^2\Omega_3^2T_2^2=K\big(K+\Omega_1^2T_2\big)$ and hence
\begin{equation}
\rss^{2}=\req^{2}\,\frac{K\big(K+\Omega_1^{2}T_2\big)}{\big(K+\Omega_1^{2}T_1\big)^{2}},
\label{eq:rsspurity_app}
\end{equation}
which is Eq.~\eqref{eq:rss-purity}. The bath enters the ratio $\rss/\req$ only through $T_1$
and $T_2$. Expanding for $\Omega_1^2\ll K$ and using $T_2-2T_1=-2\gamma_\phi T_1T_2$ together
with $K\simeq\Omega_3^2T_2$ and $\Omega_1^2/\Omega_3^2\simeq\eta^2\sin^2\theta$,
\begin{equation}
1-\frac{\rss^{2}}{\req^{2}}
\simeq-\frac{\Omega_1^{2}}{K}\big(T_2-2T_1\big)
\simeq2\eta^{2}\sin^{2}\theta\,\frac{\gamma_\phi}{\gamma(2\nth+1)},
\label{eq:purity_eta2}
\end{equation}
so the $\mathcal O(\eta^2)$ deficit is controlled by pure dephasing alone. Since
$\gamma_\phi=4\pi T_BJ'(0)$ vanishes with the bath temperature, it disappears together with
$\nth$: the two zero-temperature conditions $\nth\to0$ and $\gamma_\phi\to0$ are necessarily
taken jointly, and $T_2\to2T_1$. In that limit Eq.~\eqref{eq:rsspurity_app} factorizes
exactly as $\rss^2=\req^2\big(1-\Omega_1^4T_1^2/D^2\big)$, so with $\req\to1$ the residual
deficit is
$1-\rss^2=\Omega_1^4T_1^2/D^2\simeq\tfrac14\eta^4\sin^4\theta=\mathcal O(\eta^4)$,
which is Eq.~\eqref{eq:purity}; it vanishes only in the adiabatic limit. For the equilibrium
loop $\boldsymbol{\delta}=0$ and $r=\req\to1$ exactly.

From Eq.~\eqref{eq:eigenkets_singlephase} and the half-angle identities
$\cos^2\tfrac\theta2=\tfrac{1+\cos\theta}2$, $\sin^2\tfrac\theta2=\tfrac{1-\cos\theta}2$,
$\cos\tfrac\theta2\sin\tfrac\theta2=\tfrac{\sin\theta}2$,
\begin{equation}
|g(t)\rangle\langle g(t)|
=\frac12\begin{pmatrix}
1+\cos\theta & \sin\theta\,e^{-i\omega t}\\[2pt]
\sin\theta\,e^{+i\omega t} & 1-\cos\theta
\end{pmatrix}
=\frac12\big(\Id+\hat{\boldsymbol{n}}(t)\cdot\boldsymbol{\sigma}\big),
\label{eq:ggproj}
\end{equation}
the rank-one projector onto the instantaneous ground state. Comparing with the limit-cycle
matrix of Eq.~\eqref{eq:rhossmat} entry by entry, the populations
$(1\pm\rss\cos\tss)/2\to(1\pm\cos\theta)/2$ and the coherences
$(\rss\sin\tss/2)\,e^{\mp i(\omega t+\phi_0)}\to(\sin\theta/2)\,e^{\mp i\omega t}$ are
controlled by exactly three limits, whose rates follow from Eq.~\eqref{eq:lag}:
\begin{equation}
\begin{gathered}
\rss=\req\Big(1-\eta^2\sin^2\theta\,\gamma_\phi T_1
-\tfrac18\eta^4\sin^4\theta+\dots\Big),\\[2pt]
\tss=\theta+\frac{\delta_1}{\req}+\mathcal O(\eta^2)
=\theta-\eta\sin\theta+\mathcal O(\eta^2),\\[2pt]
\phi_0=\frac{\delta_2}{\req\sin\theta}+\dots=-\frac{\eta}{T_2\,\omega_0}+\mathcal O(\eta^2).
\end{gathered}
\label{eq:threelimits}
\end{equation}
Low temperature supplies the first limit ($\nth,\gamma_\phi/\gamma\to0$, residual
$\mathcal O(\eta^4)$); adiabaticity supplies the second and third; for the equilibrium
loop $\boldsymbol{\delta}=0$ and all three are exact. The maximal entrywise deviation is dominated by the
axis tilt, $\max_{ij}|(\rho_{\rm ss}-|g\rangle\langle g|)_{ij}|
\simeq\tfrac12\eta\sin^2\theta$, linear in $\eta$ [verified numerically: for
$d_0=d_1=1$, $\gamma=0.05$, $T_B=10^{-3}$, the deviation is $1.8\times10^{-3}$,
$8.8\times10^{-4}$, $4.4\times10^{-4}$ at $\omega=0.02,\,0.01,\,0.005$, i.e.
$0.250\,\eta$, while $\tss-\theta$ and $\phi_0$ match the rates in
Eq.~\eqref{eq:threelimits} to three digits]. The purification statement of
Sec.~\ref{sec:collapse} is thus established at the level of every matrix element, the
coherences converging to the finite value of Eq.~\eqref{eq:ggproj} rather than to zero.

\section{Open-path solid angle and geometric phases at the crossing}
\label{app:gapproofs}

This appendix derives the statements used in Sec.~\ref{sec:gapclosing}: the properties of the open-path solid angle, the closure dependence of the regularized Berry phase, and the behavior of the Uhlmann holonomy across the crossing. Uniform error estimates for the limits taken below are given in Ref.~\cite{YangNoh2026geodesic}.

\subsection{Solid angle of the open path}
\label{app:omegaC}

On the punctured sphere $S^2\setminus\{-\hat{\boldsymbol n}_0\}$, introduce polar coordinates $(\theta_n,\varphi_n)$ with $\hat{\boldsymbol n}_0$ as the north pole. Away from the antipode, the integrand of Eq.~\eqref{eq:OmegaC} becomes the monopole one-form of Eq.~\eqref{eq:berry-monopole}, 
\begin{equation}
\Omega[C]=\int_C(1-\cos\theta_n)\,\exd\varphi_n .
\label{eq:polarOmega}
\end{equation}
Because $C$ terminates at the excluded antipode, this expression is understood as an improper integral obtained by truncating the path before its endpoint.
Two properties follow. 
(i)~A great-circle arc through the antipodal pair $\pm\hat{\boldsymbol v}$ lies in a plane containing $\hat{\boldsymbol n}_0$, and its azimuth $\varphi_n$ is constant away from the antipode. The ordinary line integral along the arc therefore vanishes. This does not make the closed-loop solid angle independent of the chosen geodesic: different meridional closures approach the gauge singularity at the antipode with different azimuths and hence carry different gauge-transition contributions, as derived in Appendix~\ref{app:alphaderiv}.
(ii)~Near the crossing the transverse part of the control curve is $\boldsymbol{h}(s)=\frac12\ddot{\boldsymbol{d}}(t_0)_\perp s^2+\mathcal O(s^3)$, whose azimuth has a finite limit on each branch, so $\exd\varphi_n=\mathcal O(\exd s)$ while $1-\cos\theta_n\le2$; a window $|s|<w$ therefore contributes $\mathcal O(w)$, and the improper integral converges despite the jump in Eq.~\eqref{eq:njump}.

\subsection{Closure dependence of the regularized loop}
\label{app:alphaderiv}

Let $\hat{\boldsymbol{u}}\perp\hat{\boldsymbol{v}}$ without loss of generality; a longitudinal component of the bias only shifts the crossing time by $\mathcal O(\epsilon)$. Near the crossing the transverse part of the regularized curve is
\begin{equation}
\boldsymbol{h}_\epsilon(s)=\tfrac12\,a_\perp s^2\,\hat{\boldsymbol{a}}_\perp+\epsilon\,\hat{\boldsymbol{u}}_\perp+\mathcal O(s^3),
\qquad a_\perp=\big|\ddot{\boldsymbol{d}}(t_0)_\perp\big|,
\label{eq:hreg}
\end{equation}
so the azimuth $\varphi_n=\arg\boldsymbol{h}_\epsilon$ interpolates between two limits on the scale $s_\epsilon=(2\epsilon/a_\perp)^{1/2}$: for $|s|\gg s_\epsilon$ it approaches the azimuth of $\hat{\boldsymbol{a}}_\perp$, and for $|s|\ll s_\epsilon$ the azimuth $\alpha$ of $\hat{\boldsymbol{u}}_\perp$. The regularized loop therefore executes the azimuthal turn $\alpha$ twice, once on each side of the crossing. On the branch $s>0$ the turn occurs next to the base point, where $\theta_n\to0$ and the weight in Eq.~\eqref{eq:polarOmega} is $\mathcal O(\epsilon)$; on the branch $s<0$ it occurs next to the antipode, where $\theta_n\to\pi$ and the weight approaches $2$. The turn is thus counted asymmetrically,
\begin{equation}
\Omega(\epsilon\hat{\boldsymbol{u}})=\Omega[C]+2\alpha+\mathcal O(\epsilon),
\label{eq:alpha-derivation}
\end{equation}
which is Eq.~\eqref{eq:berryfamily} after $\gamma_B=-\Omega/2$.

\subsection{Uhlmann holonomy across the crossing}
\label{app:uhlcross}

\emph{Variable radius.} For $\rho=\frac12(\Id+r\,\hat{\boldsymbol{n}}\cdot\boldsymbol{\sigma})$ with $r(t)\in(0,1)$ and $\hat{\boldsymbol{n}}(t)$ both varying, the spectral decomposition $\rho=\lambda_+P_++\lambda_-P_-$ of Appendix~\ref{App:UhlCon} gives
\begin{equation}
\exd\sqrt\rho
=\frac{\exd\lambda_+}{2\sqrt{\lambda_+}}P_+
+\frac{\exd\lambda_-}{2\sqrt{\lambda_-}}P_-
+\big(\sqrt{\lambda_+}-\sqrt{\lambda_-}\big)\,\exd P_+ .
\label{eq:dsqrt_var}
\end{equation}
The first two terms are diagonal in the instantaneous eigenbasis and commute with $\sqrt\rho$, so they drop out of the commutator, which reduces to Eq.~\eqref{eq:comm_reduction_copy} as in the constant-$r$ case; the identity component of the transport condition, Eq.~\eqref{eq:orth_copy}, is likewise unchanged. The closed form $A_U=-\frac{i}{2}f(r(t))(\hat{\boldsymbol{n}}\times\exd\hat{\boldsymbol{n}})\cdot\boldsymbol{\sigma}$ therefore holds along any full-rank trajectory.

\emph{Uniqueness and continuity.} For $T_B>0$ the Gibbs loop is full rank everywhere, including $t_0$. With $\hat{\boldsymbol{n}}={\rm sgn}(s)\,\hat{\boldsymbol{w}}$, $\hat{\boldsymbol{w}}$ smooth, the sign cancels in $\hat{\boldsymbol{n}}\times\exd\hat{\boldsymbol{n}}$, and $f(r(t))$ is smooth with $f(r(t_0))=0$; the connection is a smooth, bounded $\mathfrak u(2)$-valued one-form on the whole loop, its path-ordered exponential $U_C$ exists, and it depends continuously on $\beta$. Since $A_U$ is traceless and anti-Hermitian, $U_C\in SU(2)$ and $\mathrm{Tr}[\frac12\Id\,U_C]$ is real, so a loop based at $t_0$ carries no phase information; we therefore choose and fix a reference time $t_r\neq t_0$, and $\gamma_U=\Argop\operatorname{Tr}[\rho(t_r)U_C(t_r)]$ is continuous in $T_B$ wherever the trace is nonzero.

\emph{Pure-state limit.} Split the loop at $|s|=w$. Outside the window, $d\geq c_w>0$, so $f\to1$ uniformly as $T_B\to0$ and $A_U$ converges to the pure-state parallel-transport generator, whose accumulated phase is the line integral of Eq.~\eqref{eq:polarOmega} outside the window; inside the window, $\|A_U\|\le\frac12f\,|\hat{\boldsymbol{w}}\times\exd\hat{\boldsymbol{w}}|$ with $f\le1$ bounds the contribution to $U_C$ by $\mathcal O(w)$ uniformly in $T_B$, and the contribution to $\Omega[C]$ is $\mathcal O(w)$ by Appendix~\ref{app:omegaC}. Taking $T_B\to0$ and then $w\to0$ yields Eq.~\eqref{eq:uhlmann-unique}.

\emph{Reflection-symmetric crossing.} If a plane $\Pi$ through the origin contains $\boldsymbol{d}(t)$ for all $t$, then $\hat{\boldsymbol{w}}$ is confined to the great circle $\Pi\cap S^2$ and $\hat{\boldsymbol{w}}\times\exd\hat{\boldsymbol{w}}=\hat{\boldsymbol{u}}\,\exd\varphi_n$, with $\hat{\boldsymbol{u}}$ the mirror normal and $\varphi_n$ the angle along the circle. All values of $A_U$ then commute, and
\begin{equation}
U_C=\exp\!\Big[-\frac{i}{2}\,\Phi_U\,\hat{\boldsymbol{u}}\cdot\boldsymbol{\sigma}\Big],
\qquad
\Phi_U=\oint f\big(r(t)\big)\,\exd\varphi_n .
\label{eq:UC_reflection}
\end{equation}
With the base point $t_r\neq t_0$, the axis $\hat{\boldsymbol{n}}(t_r)\in\Pi$ is orthogonal to $\hat{\boldsymbol{u}}$, so
\begin{equation}
\mathrm{Tr}\big[\rho(t_r)\,U_C\big]=\cos(\Phi_U/2),
\label{eq:trace_reflection}
\end{equation}
which is real, and $\gamma_U\in\{0,\pi\}$. If the direction winds monotonically along the circle---the generic case of a simple crossing, with net advance $\oint\exd\varphi_n=\pi$---then $|\Phi_U|<\pi$ strictly, because $f<1$ for $T_B>0$ and $f=0$ at the crossing; hence
\begin{equation}
\gamma_U=0\qquad\text{for every }T_B>0 .
\label{eq:gammaU_reflection}
\end{equation}
As $T_B\to0$, $\Phi_U\to\pi^-$: the interference amplitude $\cos(\Phi_U/2)$ vanishes while its phase stays pinned at $0$, consistent with the normal-bias Berry values $\mp\pi/2$, whose contributions cancel pairwise. For planar drives that backtrack or carry additional winding, $|\Phi_U|$ can cross $\pi$ at a finite temperature, at which $\gamma_U$ jumps to $\pi$, still without regularization ambiguity.

\bibliography{References_Uhlmann}

@article{Berry1984,
  author  = {Berry, M. V.},
  title   = {Quantal phase factors accompanying adiabatic changes},
  journal = {Proc. R. Soc. Lond. A},
  volume  = {392},
  number  = {1802},
  pages   = {45--57},
  year    = {1984},
  doi     = {10.1098/rspa.1984.0023}
}

@article{AharonovAnandan1987,
  author  = {Aharonov, Y. and Anandan, J.},
  title   = {Phase Change during a Cyclic Quantum Evolution},
  journal = {Phys. Rev. Lett.},
  volume  = {58},
  number  = {16},
  pages   = {1593--1596},
  year    = {1987},
  doi     = {10.1103/PhysRevLett.58.1593}
}

@article{TomitaChiao1986,
  author  = {Tomita, Akira and Chiao, Raymond Y.},
  title   = {Observation of {Berry's} Topological Phase by Use of an
             Optical Fiber},
  journal = {Phys. Rev. Lett.},
  volume  = {57},
  number  = {8},
  pages   = {937--940},
  year    = {1986},
  doi     = {10.1103/PhysRevLett.57.937}
}

@article{BitterDubbers1987,
  author  = {Bitter, T. and Dubbers, D.},
  title   = {Manifestation of {Berry's} topological phase in neutron spin
             rotation},
  journal = {Phys. Rev. Lett.},
  volume  = {59},
  number  = {3},
  pages   = {251--254},
  year    = {1987},
  doi     = {10.1103/PhysRevLett.59.251}
}

@article{Tonomura1988new,
  title={New results on the Aharonov-Bohm effect with electron interferometry},
  author={Tonomura, Akira},
  journal={Physica B+ C},
  volume={151},
  number={1-2},
  pages={206--213},
  year={1988},
  publisher={Elsevier}
}

@article{Tewari1989,
  author  = {Tewari, Surya P.},
  title   = {Berry's phase in a two-level atom},
  journal = {Phys. Rev. A},
  volume  = {39},
  number  = {11},
  pages   = {6082--6085},
  year    = {1989},
  doi     = {10.1103/PhysRevA.39.6082}
}

@article{SjoqvistGPQI2015,
  author  = {Sj{\"o}qvist, Erik},
  title   = {Geometric phases in quantum information},
  journal = {Int. J. Quantum Chem.},
  volume  = {115},
  number  = {19},
  pages   = {1311--1326},
  year    = {2015},
  doi     = {10.1002/qua.24941}
}

@article{Uhlmann1986,
  author  = {Uhlmann, Armin},
  title   = {Parallel transport and ``quantum holonomy'' along density
             operators},
  journal = {Rep. Math. Phys.},
  volume  = {24},
  number  = {2},
  pages   = {229--240},
  year    = {1986},
  doi     = {10.1016/0034-4877(86)90055-8}
}

@article{Hubner1993,
  author  = {H{\"u}bner, Matthias},
  title   = {Computation of {Uhlmann's} parallel transport for density
             matrices and the {Bures} metric on three-dimensional
             {Hilbert} space},
  journal = {Phys. Lett. A},
  volume  = {179},
  number  = {4-5},
  pages   = {226--230},
  year    = {1993},
  doi     = {10.1016/0375-9601(93)90668-P}
}

@article{Sjoqvist2000,
  author  = {Sj{\"o}qvist, E. and Pati, A. K. and Ekert, A. and
             Anandan, J. S. and Ericsson, M. and Oi, D. K. L. and
             Vedral, V.},
  title   = {Geometric Phases for Mixed States in Interferometry},
  journal = {Phys. Rev. Lett.},
  volume  = {85},
  number  = {14},
  pages   = {2845--2849},
  year    = {2000},
  doi     = {10.1103/PhysRevLett.85.2845}
}

@article{Tidstrom2003,
  author  = {Tidstr{\"o}m, Jonas and Sj{\"o}qvist, Erik},
  title   = {Uhlmann's geometric phase in presence of isotropic
             decoherence},
  journal = {Phys. Rev. A},
  volume  = {67},
  number  = {3},
  pages   = {032110},
  year    = {2003},
  doi     = {10.1103/PhysRevA.67.032110}
}

@article{Guo2020,
  author  = {Guo, Hao and Hou, Xu-Yang and He, Yan and Chien, Chih-Chun},
  title   = {Dynamic process and {Uhlmann} process: Incompatibility and
             dynamic phase of mixed quantum states},
  journal = {Phys. Rev. B},
  volume  = {101},
  number  = {10},
  pages   = {104310},
  year    = {2020},
  doi     = {10.1103/PhysRevB.101.104310}
}

@article{Guo2020erratum,
  author  = {Guo, Hao and Hou, Xu-Yang and He, Yan and Chien, Chih-Chun},
  title   = {Erratum: Dynamic process and {Uhlmann} process:
             Incompatibility and dynamic phase of mixed quantum states
             [Phys. Rev. B {\bf 101}, 104310 (2020)]},
  journal = {Phys. Rev. B},
  volume  = {102},
  number  = {13},
  pages   = {139901(E)},
  year    = {2020},
  doi     = {10.1103/PhysRevB.102.139901}
}

@article{WangUBC2023,
  author  = {Wang, Xin and Hou, Xu-Yang and Zhou, Zheng and Guo, Hao and
             Chien, Chih-Chun},
  title   = {Uhlmann phase of coherent states and the {Uhlmann-Berry}
             correspondence},
  journal = {SciPost Phys. Core},
  volume  = {6},
  number  = {1},
  pages   = {024},
  year    = {2023},
  doi     = {10.21468/SciPostPhysCore.6.1.024}
}

@article{Hou2023comparative,
  author  = {Hou, Xu-Yang and Wang, Xin and Zhou, Zheng and Guo, Hao and
             Chien, Chih-Chun},
  title   = {Geometric phases of mixed quantum states: A comparative study
             of interferometric and {Uhlmann} phases},
  journal = {Phys. Rev. B},
  volume  = {107},
  number  = {16},
  pages   = {165415},
  year    = {2023},
  doi     = {10.1103/PhysRevB.107.165415}
}

@article{Hou2024QGT,
  author  = {Hou, Xu-Yang and Zhou, Zheng and Wang, Xin and Guo, Hao and
             Chien, Chih-Chun},
  title   = {Local geometry and quantum geometric tensor of mixed states},
  journal = {Phys. Rev. B},
  volume  = {110},
  number  = {3},
  pages   = {035144},
  year    = {2024},
  doi     = {10.1103/PhysRevB.110.035144}
}

@article{He2018ThermalChern,
  author  = {He, Yan and Guo, Hao and Chien, Chih-Chun},
  title   = {Thermal {Uhlmann-Chern} number from the {Uhlmann} connection
             for extracting topological properties of mixed states},
  journal = {Phys. Rev. B},
  volume  = {97},
  number  = {23},
  pages   = {235141},
  year    = {2018},
  doi     = {10.1103/PhysRevB.97.235141}
}

@article{Hou2021spinj,
  author  = {Hou, Xu-Yang and Guo, Hao and Chien, Chih-Chun},
  title   = {Finite-temperature topological phase transitions of spin-$j$
             systems in {Uhlmann} processes: General formalism and
             experimental protocols},
  journal = {Phys. Rev. A},
  volume  = {104},
  number  = {2},
  pages   = {023303},
  year    = {2021},
  doi     = {10.1103/PhysRevA.104.023303}
}

@article{Villavicencio2023,
  author  = {Villavicencio, J. and Cota, E. and Rojas, F. and
             Maytorena, Jes{\'u}s A. and Morachis Galindo, D. and
             Nieto-Guadarrama, F.},
  title   = {Thermal {Uhlmann} phase in a locally driven two-spin system},
  journal = {Phys. Rev. A},
  volume  = {107},
  number  = {6},
  pages   = {062222},
  year    = {2023},
  doi     = {10.1103/PhysRevA.107.062222}
}

@article{Wang2025spin,
  author  = {Wang, Xin and Tang, Jia-Chen and Hou, Xu-Yang and Guo, Hao and
             Chien, Chih-Chun},
  title   = {Mixed-state geometric phases of coherent and squeezed spin
             states},
  journal = {Phys. Rev. B},
  volume  = {111},
  number  = {23},
  pages   = {235450},
  year    = {2025},
  doi     = {10.1103/98sq-16bz}
}

@article{Carollo2018,
  author  = {Carollo, Angelo and Spagnolo, Bernardo and Valenti, Davide},
  title   = {Uhlmann curvature in dissipative phase transitions},
  journal = {Sci. Rep.},
  volume  = {8},
  pages   = {9852},
  year    = {2018},
  doi     = {10.1038/s41598-018-27362-9}
}

@article{He2022Lindblad,
  author  = {He, Yan and Chien, Chih-Chun},
  title   = {Uhlmann holonomy against {Lindblad} dynamics of topological
             systems at finite temperatures},
  journal = {Phys. Rev. B},
  volume  = {106},
  number  = {2},
  pages   = {024310},
  year    = {2022},
  doi     = {10.1103/PhysRevB.106.024310}
}

@article{Viotti2023,
  author  = {Viotti, Ludmila and Gramajo, Ana Laura and Villar, Paula I. and
             Lombardo, Fernando C. and Fazio, Rosario},
  title   = {Geometric phases along quantum trajectories},
  journal = {Quantum},
  volume  = {7},
  pages   = {1029},
  year    = {2023},
  doi     = {10.22331/q-2023-06-13-1029}
}

@article{Rivas2013,
  author  = {Rivas, A. and Viyuela, O. and Martin-Delgado, M. A.},
  title   = {Density-matrix {Chern} insulators: Finite-temperature
             generalization of topological insulators},
  journal = {Phys. Rev. B},
  volume  = {88},
  number  = {15},
  pages   = {155141},
  year    = {2013},
  doi     = {10.1103/PhysRevB.88.155141}
}

@article{Viyuela2014_2D,
  author  = {Viyuela, O. and Rivas, A. and Martin-Delgado, M. A.},
  title   = {Two-Dimensional Density-Matrix Topological Fermionic Phases:
             Topological {Uhlmann} Numbers},
  journal = {Phys. Rev. Lett.},
  volume  = {113},
  number  = {7},
  pages   = {076408},
  year    = {2014},
  doi     = {10.1103/PhysRevLett.113.076408}
}

@article{Viyuela2015SPT,
  author  = {Viyuela, O. and Rivas, A. and Martin-Delgado, M. A.},
  title   = {Symmetry-protected topological phases at finite temperature},
  journal = {2D Mater.},
  volume  = {2},
  number  = {3},
  pages   = {034006},
  year    = {2015},
  doi     = {10.1088/2053-1583/2/3/034006}
}

@article{Viyuela2018Obs,
  author  = {Viyuela, O. and Rivas, A. and Gasparinetti, S. and
             Wallraff, A. and Filipp, S. and Martin-Delgado, M. A.},
  title   = {Observation of topological {Uhlmann} phases with
             superconducting qubits},
  journal = {npj Quantum Inf.},
  volume  = {4},
  pages   = {10},
  year    = {2018},
  doi     = {10.1038/s41534-017-0056-9}
}

@article{Andersson2016,
  author  = {Andersson, Ole and Bengtsson, Ingemar and Ericsson, Marie and
             Sj{\"o}qvist, Erik},
  title   = {Geometric phases for mixed states of the {Kitaev} chain},
  journal = {Phil. Trans. R. Soc. A},
  volume  = {374},
  number  = {2069},
  pages   = {20150231},
  year    = {2016},
  doi     = {10.1098/rsta.2015.0231}
}

@article{vanCaspel2019,
  author  = {van Caspel, Moos T. and Tapias Arze, Sergio Enrique and
             P{\'e}rez Castillo, Isaac},
  title   = {Dynamical signatures of topological order in the
             driven-dissipative {Kitaev} chain},
  journal = {SciPost Phys.},
  volume  = {6},
  number  = {2},
  pages   = {026},
  year    = {2019},
  doi     = {10.21468/SciPostPhys.6.2.026}
}

@misc{YangNoh2026geodesic,
  author = {Yang, Hyeonseok and Noh, Changsuk},
  title  = {Geometric phase of open paths and a geodesic-selection rule at a level degeneracy},
  year   = {2026},
  note   = {arXiv:2608.19679}
}

@article{BerryWilkinson1984,
  author  = {Berry, M. V. and Wilkinson, M.},
  title   = {Diabolical points in the spectra of triangles},
  journal = {Proc. R. Soc. Lond. A},
  volume  = {392},
  pages   = {15},
  year    = {1984}
}

@article{Dirac1931,
  author  = {Dirac, P. A. M.},
  title   = {Quantised singularities in the electromagnetic field},
  journal = {Proc. R. Soc. Lond. A},
  volume  = {133},
  pages   = {60},
  year    = {1931}
}

@article{Simon1983,
  author  = {Simon, Barry},
  title   = {Holonomy, the quantum adiabatic theorem, and {B}erry's phase},
  journal = {Phys. Rev. Lett.},
  volume  = {51},
  pages   = {2167},
  year    = {1983}
}

@article{Pancharatnam1956,
  author  = {Pancharatnam, S.},
  title   = {Generalized theory of interference, and its applications},
  journal = {Proc. Indian Acad. Sci. A},
  volume  = {44},
  pages   = {247},
  year    = {1956}
}

@article{SamuelBhandari1988,
  author  = {Samuel, J. and Bhandari, R.},
  title   = {General setting for {B}erry's phase},
  journal = {Phys. Rev. Lett.},
  volume  = {60},
  pages   = {2339},
  year    = {1988}
}

@article{MukundaSimon1993,
  author  = {Mukunda, N. and Simon, R.},
  title   = {Quantum kinematic approach to the geometric phase. {I}. {G}eneral formalism},
  journal = {Ann. Phys. (N.Y.)},
  volume  = {228},
  pages   = {205},
  year    = {1993}
}

@article{RakhechaWagh1996,
  author  = {Rakhecha, V. C. and Wagh, A. G.},
  title   = {Geometric phase \`a la {P}ancharatnam},
  journal = {Pramana -- J. Phys.},
  volume  = {46},
  pages   = {315},
  year    = {1996}
}

@article{Wagh1998,
  author  = {Wagh, A. G. and Rakhecha, V. C. and Fischer, P. and Ioffe, A.},
  title   = {Neutron interferometric observation of noncyclic phase},
  journal = {Phys. Rev. Lett.},
  volume  = {81},
  pages   = {1992},
  year    = {1998}
}

@article{Zhou2020,
  author  = {Zhou, Zhifan and Margalit, Yair and Moukouri, Samuel and Meir, Yigal and Folman, Ron},
  title   = {An experimental test of the geodesic rule proposition for the noncyclic geometric phase},
  journal = {Sci. Adv.},
  volume  = {6},
  pages   = {eaay8345},
  year    = {2020}
}

@article{GarzaSoto2023,
  author  = {Garza-Soto, Luis and Hagen, Nathan and Lopez-Mago, Dorilian},
  title   = {Deciphering {P}ancharatnam's discovery of geometric phase: retrospective},
  journal = {J. Opt. Soc. Am. A},
  volume  = {40},
  pages   = {925},
  year    = {2023}
}

@article{Viyuela2014_1D,
  author  = {Viyuela, O. and Rivas, A. and Martin-Delgado, M. A.},
  title   = {Uhlmann phase as a topological measure for one-dimensional fermion systems},
  journal = {Phys. Rev. Lett.},
  volume  = {112},
  pages   = {130401},
  year    = {2014}
}

@article{Wang2025degenerate,
  author  = {Wang, Xin and Hou, Xu-Yang and Guo, Hao and Chien, Chih-Chun},
  title   = {Uhlmann and scalar {W}ilczek--{Z}ee phases of degenerate quantum systems},
  journal = {Phys. Rev. B},
  volume  = {112},
  pages   = {134315},
  year    = {2025}
}

@article{Landau1932,
  author  = {Landau, L. D.},
  title   = {Zur {T}heorie der {E}nergie\"ubertragung. {II}},
  journal = {Phys. Z. Sowjetunion},
  volume  = {2},
  pages   = {46},
  year    = {1932}
}

@article{Zener1932,
  author  = {Zener, C.},
  title   = {Non-adiabatic crossing of energy levels},
  journal = {Proc. R. Soc. Lond. A},
  volume  = {137},
  pages   = {696},
  year    = {1932}
}

@article{AoRammer1989,
  author  = {Ao, P. and Rammer, J.},
  title   = {Influence of dissipation on the {L}andau-{Z}ener transition},
  journal = {Phys. Rev. Lett.},
  volume  = {62},
  pages   = {3004},
  year    = {1989}
}

@article{Wubs2006,
  author  = {Wubs, M. and Saito, K. and Kohler, S. and H\"anggi, P. and Kayanuma, Y.},
  title   = {Gauging a quantum heat bath with dissipative {L}andau-{Z}ener transitions},
  journal = {Phys. Rev. Lett.},
  volume  = {97},
  pages   = {200404},
  year    = {2006}
}

\end{document}